\documentclass[12pt]{article}

\usepackage{amsthm}
\usepackage{amsmath}
\usepackage{amssymb}
\usepackage{natbib}
\usepackage{url}  %
\usepackage{graphicx}
\usepackage{bm}
\usepackage{floatrow}
\usepackage{silence} 
\usepackage{caption}
\usepackage{subcaption}
\usepackage[title]{appendix}
\usepackage[shortlabels]{enumitem}
\usepackage{booktabs} %
\usepackage{longtable} %
\usepackage{ulem} %
\usepackage{tikz}
\usepackage{siunitx}
\usepackage{changepage}
\usepackage[colorlinks,citecolor=blue,urlcolor=blue,filecolor=blue]{hyperref}
\usepackage[ruled]{algorithm2e}
\usepackage{tabularx}
\usepackage{geometry}
\newtheorem{theorem}{Theorem}
\newtheorem{definition}{Definition}

\newcommand{\baselineitem}{\mu}
\newcommand{\baselinepart}{\bm{\baselineitem}}
\newcommand{\shrinkitem}{\omega}
\newcommand{\shrink}{\bm{\shrinkitem}}
\newcommand{\partitionitem}{\pi}
\newcommand{\partition}{\bm{\partitionitem}}
\newcommand{\partitionupto}[1]{\partition_{\permitem_{1:#1}}}
\newcommand{\permitem}{\sigma}
\newcommand{\perm}{\bm{\permitem}}
\newcommand{\baselinedist}{p_\text{b}}
\newcommand{\shrinkdist}{p_\text{sp}}
\newcommand{\condprob}{\text{Pr}}

\newcommand{\prb}{\ensuremath{\text{Pr}_\text{b}}}

\newcommand{\numi}{n} %

\begin{document}

\def\spacingset#1{\renewcommand{\baselinestretch}%
{#1}\small\normalsize} \spacingset{1}

\newcommand{\thankstext}{The authors thank the editor, associate editor, and two reviewers
 for their time and constructive comments which significantly improved the
 manuscript.  The authors report no conflicts of interest.}

  \title{\bf \vspace{-2ex}Dependent Random Partitions by\\ Shrinking Toward an Anchor}
  \author{David B.\ Dahl\thanks{\thankstext{}} \thanks{ORCID: 0000-0002-8173-1547}\\
    Department of Statistics, Brigham Young University\vspace{1.5ex}\\
    Richard L.\ Warr\thanks{ORCID: 0000-0001-8508-3105}\\
    Department of Statistics, Brigham Young University\vspace{1.5ex}\\
    Thomas P.\ Jensen\thanks{ORCID: 0009-0000-0015-3881}\\
    Berry Consultants LLC}
  \maketitle

\newcommand{\dbd}[1]{\textcolor{red}{#1}}
\newcommand{\rlw}[1]{\textcolor{green}{#1}}
\newcommand{\tpj}[1]{\textcolor{orange}{#1}}

\vspace{-4ex}
\begin{abstract}
Although exchangeable processes from Bayesian nonparametrics have been used as a generating mechanism for random partition models, we deviate from this paradigm to explicitly incorporate clustering information in the formulation of our random partition model. Our shrinkage partition distribution takes any partition distribution and shrinks its probability mass toward an anchor partition. We show how this provides a framework to model hierarchically-dependent and temporally-dependent random partitions. The shrinkage parameter controls the degree of dependence, accommodating at its extremes both independence and complete equality. Since \textit{a priori} knowledge of items may vary, our formulation allows the degree of shrinkage toward the anchor to be item-specific. Our random partition model has a tractable normalizing constant which allows for standard Markov chain Monte Carlo algorithms for posterior sampling. We prove intuitive theoretical properties for our distribution and compare it to related partition distributions. We show that our model provides better out-of-sample fit in a real data application.
\end{abstract}

\noindent%
\vfill

\newpage
\spacingset{1.7} %

\section{Introduction}
Random partition models are flexible Bayesian prior distributions which
accommodate heterogeneity and the borrowing of strength by postulating that
data or parameters are generated from latent clusters.
Exchangeable random partition models arise from Bayesian nonparametric (BNP)
models, which are very flexible and often involve rich modeling techniques
applied to specific problems.
In the presence of additional information that ought to influence
the partition, the exchangeability constraint is untenable and
researchers have developed nonexchangeable random partition models that
are \textit{not} obtained by marginalizing over a random measure $G$.
Examples include \cite{muller2011product}, \cite{blei2011distance},
\cite{airoldi2014generalized}, and \cite{dahl2017random} which use covariates or distances to influence a partition distribution.
Some have sought partition distributions which directly incorporate prior
knowledge on the partition itself. Rather than incorporating covariates or
distances in a partition distribution, one might wish to use a ``best guess'' as
to the value of the partition, yet may not be completely certain and therefore
may be unwilling to fix the partition on this value.  Instead, a modeler may
wish to use a partition distribution that is anchored on this ``best guess'' partition yet allow for deviations from it.

We introduce a partition distribution with a \textit{shrinkage}
parameter $\shrink$ that governs the concentration of probability mass between two
competing elements: a \textit{baseline}
partition distribution $\baselinedist$ and an \textit{anchor} partition
$\baselinepart$.  We call our distribution the shrinkage partition (SP) distribution.  Our approach builds on the pioneering work
of \citet{smith2020demand}'s location scale partition (LSP) and \citet{paganin2020centered}'s
centered partition process (CPP). 
Our SP distribution and both the LSP and CPP distributions
are influenced by an anchor partition (called the
``location'' partition and the ``centered'' partition in their respective papers).
However, the LSP depends on the arbitrary
ordering of the data and the distribution of deviations from the anchor
partition is immutable and intrinsically embedded, which limits modeling
flexibility. Both our SP distribution and the CPP
allow for any baseline partition distribution,
but the probability mass function (pmf) of the CPP is only specified up to the
normalizing constant.  The nature of the combinatorics makes the normalizing
constant intractable, so the shrinkage parameter must be fixed using a
computationally intensive cross-validation procedure and posterior inference on hyperparameters (e.g., the shrinkage parameter)
is not practically feasible.

In contrast, our SP distribution has several desirable properties.  First, the
SP has a tractable normalizing constant, so standard Markov chain Monte Carlo
(MCMC) techniques can be used for posterior inference on hyperparameters.
Second, the SP does not depend on
the order in which data are observed, making the data analysis invariant to what
may be an arbitrary ordering.  Third, the SP allows for any baseline partition
distribution to govern the distribution of partitions informed by the anchor
partition.
Fourth, prior knowledge about the clustering of items may be
different across the items and our SP is unique in allowing differential shrinkage
toward the anchor.

Further, whereas the LSP and CPP were introduced as prior distributions for
a single partition in situations where the modeler has a prior guess as to
the value of the partition, we note that our formulation --- in addition to
being suitable for this case --- also permits building models for dependent
random partitions. \cite{page2022dependent} was the first to directly model
temporally-dependent partitions that are evenly spaced in time. Our approach adds flexibility,
permitting temporally-dependent random partition models that are not
necessarily observed on a uniformly-spaced grid. Further, our SP permits not
only temporal dependence, but other forms of dependence.
\newcommand{\firstnovel}{%
\cite{Camerlenghi2019distribution} and \cite{argiento2020hierarchical} show how to build hierarchical clustering models, however, as demonstrated in \cite{page2022dependent} dependence induced between partitions via random measures is somewhat limited.  \citet{page2022dependent} showed that to achieve the full spectrum of dependence, from independence to equality, dependence must be induced directly on the partitions.
We believe
that our SP is the first to allow hierarchically-dependent partitions induced directly on the partitions.}
\firstnovel{}
\cite{song2023clustering} recently used our SP  distribution as a basis for
their innovative hierarchical bi-clustering model of mouse-tracking data.
In contemporaneous work, \cite{paganin2023informed} and \cite{dombowsky2023product} are two recent
pre-prints on dependent random partitions.

The SP distribution scales well in the number of items being clustered.  It
has properties that one would expect in a partition distribution, such
as the ability to control the distribution of the number of subsets, influence
the distribution of cluster sizes, and generally behaves like other partition
distributions in the literature, with the added feature of shrinking toward an
anchor partition.  Software implementing our SP distribution is available as an
R package based on Rust (\url{https://github.com/dbdahl/gourd-package}).

The remainder of the paper is organized as follows.  In Section
\ref{sec:existing}, we review the LSP and CPP and put them in common notation
for ease in understanding the novelty of our SP distribution, which is
detailed in Section \ref{sec:spd}.  Properties of the SP distribution are
detailed in Section \ref{sec:prior_exploration} and models for dependent partition
models based on the SP distribution are discussed in Section \ref{sec:dependRPs}.  An empirical study in Section
\ref{sec:AppRTE} shows that our SP distribution, based on 10-fold cross
validation, compares favorably to the LSP and CPP for a single partition and
that our SP can be used for hierarchically-dependent and temporally-dependent
partitions to improve performance beyond what is possible with models
that ignore dependence among partitions.

\section{Existing Distributions Indexed by a Partition}
\label{sec:existing}

Our shrinkage partition distribution itself was inspired by
\citet{paganin2020centered}'s CPP and \citet{smith2020demand}'s LSP. A key
parameter in all three distributions is what we term the anchor partition $
\baselinepart$, a partition of the integers $\{1,\dots,n\}$ which represents
the \textit{a priori} estimate of the population partition. The anchor partition
plays the same role as the ``centered'' partition in the CPP and the ``location''
partition in the LSP.  As noted by \citet{dahl2021discussion}, words like
``centered'' or ``location'' may be misnomers. We use the term ``anchor'' to suggest
that the probability distribution is ``tethered to'' --- but not ``centered
on'' or ``located at'' --- the anchor partition $\baselinepart$.  How closely the partition distribution reflects 
$\baselinepart$ is a function of what we term the shrinkage parameter $\shrink$.

In this paper
the anchor partition $\baselinepart$ is expressed in terms of cluster
labels $\baselinepart = (\baselineitem_1,\ldots,\baselineitem_n)$ such that two
items have the same
label (i.e.,  $\baselineitem_i = \baselineitem_j$) if and
only if they belong to the same subset in the anchor partition.   Likewise, we
use $\partition$ to denote the vector of cluster labels of a random partition.
Cluster labels are assumed to be in canonical form, i.e., cluster label
``1'' is used for the first item (and all items clustered with it), cluster
label ``2'' is used for the next item (and all items clustered with it) that is not
clustered with the first item, etc.
Studying the pmf of each distribution helps one see the similarities with our SP 
distribution, and what motivated us to develop our new partition distribution.
For the sake of comparison, we express each distribution in common notation.

\subsection{Centered Partition Process}

The CPP of \citet{paganin2020centered} is formed from a baseline partition distribution $\baselinedist(\partition)$ and a distance function $d(\partition, \baselinepart)$ which 
measures the discrepancy between two partitions.  The pmf of the CPP is
$p_\text{cp}(\partition \mid \baselinepart,\shrinkitem,d,\baselinedist) \ \propto \ 
\baselinedist(\partition) \, \exp{(-\shrinkitem \, d(\partition,\baselinepart))},$
where $\shrinkitem \ge 0$ is a univariate shrinkage
parameter controlling the magnitude of the penalty for discrepancy with the
anchor partition $\baselinepart$. 
The baseline partition distribution
$\baselinedist(\partition)$ will likely depend on other parameters (e.g.,
a concentration parameter) but we suppress them here for simplicity. Note that $p_\text{cp}(\partition \mid 
\baselinepart,\shrinkitem,d,\baselinedist)$ reduces to
the baseline partition distribution $\baselinedist(\partition)$ as $\shrinkitem \rightarrow 0$. 
Conversely, as $\shrinkitem \rightarrow \infty$, all probability concentrates on the anchor 
partition $\baselinepart$, i.e. $p_\text{cp}
(\partition = \baselinepart \mid \baselinepart,\shrinkitem,d,\baselinedist) \rightarrow 1$. 
While any distance function $d(\partition, \baselinepart)$ for partitions could be used,
\citet{paganin2020centered} focus on the variation of information
\citep{meilua2007comparing,wade2018bayesian}.
We use the notation 
$\partition \sim \text{CPP}(\baselinepart,\shrinkitem, d, \baselinedist)$ for a partition
whose distribution is the CPP with anchor partition $\baselinepart$, shrinkage parameter $\shrinkitem$, distance function $d$, and baseline distribution $\baselinedist$.

A key challenge with the CPP is that the
pmf is only specified up to proportionality.
While the normalizing constant can theoretically be obtained through enumeration,
the number of possible partitions of $n$
items (as given by the Bell number) quickly makes enumeration
infeasible beyond 20 or so items.
The lack of a normalizing constant precludes posterior inference on other parameters associated with the baseline distribution (e.g., a concentration parameter).
Further, as a practical matter, the shrinkage parameter $\shrinkitem$ must be fixed for data analysis, making the CPP unsuitable for modeling dependent partitions.

\subsection{Location-Scale Partition Distribution} \label{sec:lsp}
As the name implies, \citet{smith2020demand} originally
expressed their location-scale partition (LSP)
distribution using a scale parameter.
We reparameterize the LSP's pmf, for the sake of comparison, to use a univariate shrinkage
parameter $\shrinkitem$, which is the reciprocal of their original scale parameter. The
LSP has a constructive definition, in which a partition $\partition$ is
obtained by sequentially allocating items.
There may be common values among $\partitionitem_1,\ldots,\partitionitem_i$ and we
let $\{\partitionitem_1,\ldots,\partitionitem_i \}$ denote the set of unique cluster labels 
after the $i^{\text{th}}$ item is allocated and, therefore, $\left| \{\partitionitem_1,\ldots,\partitionitem_i \} 
\right|$ is the cardinality of that set or, in other words, the number of clusters in the 
partition after item $i$ is allocated.
Note that, for the LSP, the order in which items are allocated affects the probability of a 
partition, making data analysis subject to what is often an arbitrary ordering.  To overcome this deficiency and to make the comparisons in Section \ref{sec:AppRTE} more fair, we generalize the original LSP
using a technique of \citet{dahl2017random}.  Specifically, items are allocated in an order given by a permutation $\perm = (\permitem_1, \dots,
\permitem_n)$ of the integers $\{1,\dots,n\}$, where the \textit{k}$^\text{th}$ item
allocated is the $\permitem_k^\text{ th}$ item in the model or dataset, and a uniform prior is placed on the permutation $\perm$.  Thus the LSP's modified pmf is:
\begin{equation*}
p_\text{\,lsp}(\partition \mid \baselinepart, \shrinkitem,\perm) = \prod_{k=1}^n 
\condprob_\text{\,lsp}\left(\partitionitem_{\permitem_k} = c \mid 
\partitionitem_{\permitem_1},\ldots,\partitionitem_{\permitem_{k-1}},\baselinepart, \shrinkitem,\perm 
\right),
\end{equation*}
where by definition $\condprob_\text{\,lsp}\left( \partitionitem_{\permitem_1} = 1 \mid 
 \baselinepart, \shrinkitem, \perm \right) = 1$ and for $k = 2,\ldots,n$:
\begin{align}
\label{eq_LSP_pmf_cases}
\condprob_\text{\,lsp} & \left(\partitionitem_{\permitem_k} = c \mid 
\partitionitem_{\permitem_1},\ldots,\partitionitem_{\permitem_{k-1}},\baselinepart, 
\shrinkitem,\perm \right) \propto \\ &
\begin{cases}
\displaystyle \frac{1 + \shrinkitem \sum_{j=1}^{k-1} \text{I} \left\{ \partitionitem_{\permitem_j}=c 
\right\} \text{I}\left\{\baselineitem_{\permitem_j} = \baselineitem_{\permitem_k} \right\} }{1 + 
\left| \{\baselineitem_{\permitem_1},\ldots,\baselineitem_{\permitem_{k-1}} \} \right| +\shrinkitem 
\sum_{j=1}^{k-1} \text{I} \left\{ \partitionitem_{\permitem_j}=c \right\} } & \text{for } c \in 
\{\partitionitem_{\permitem_1},\ldots,\partitionitem_{\permitem_{k-1}}\} \\
\vspace{-0.75ex} & \vspace{-0.75ex} \\
\displaystyle\frac{1 + \shrinkitem \, \text{I}\left\{ \sum_{j=1}^{k-1} \text{I} \{ 
\baselineitem_{\permitem_j} = \baselineitem_{\permitem_k} \} =0 \right\}
}{1 + \left| \{\baselineitem_{\permitem_1},\ldots,\baselineitem_{\permitem_{k-1}} \} \right| + 
\shrinkitem} & \text{for } c = 
\left|\{\partitionitem_{\permitem_1},\ldots,\partitionitem_{\permitem_{k-1}}\} \right|+1.
\end{cases} \nonumber
\end{align}
Following \cite{dahl2017random}, the key to obtaining a normalizing constant for the LSP and our SP distribution comes from sequential allocation.

We use the notation 
$\partition \sim \text{LSP}(\baselinepart,\shrinkitem)$ for a partition
whose distribution is our modification of the LSP, marginalizing over $\perm$.
We note that using a permutation parameter $\perm$ (as introduced by \citealt{dahl2017random}) and marginalizing over it makes the LSP invariant to the ordering of the data.  However, this invariance to the ordering of the data does not imply the revised LSP (nor our SP distribution) is exchangeable since the probability of a partition under those distributions is not solely a function of cluster sizes.  See \citet[pg. 438]{BnpBookGhosal} for the definition of exchangeable random partitions.
As with the CPP, $p_\text{\,lsp}(\partition = 
\baselinepart \mid \baselinepart,\shrinkitem) \rightarrow 1$ as $\omega \rightarrow \infty$.

Noticeably absent from the LSP are parameters to control the distribution beyond the 
anchor partition $\baselinepart$ and the shrinkage $\shrinkitem$.
As $\shrinkitem \rightarrow 0$, the sequential allocation probabilities found in 
(\ref{eq_LSP_pmf_cases}) reduce to a uniform selection among existing clusters and a new 
cluster.  There is no mechanism to control the number of clusters.
The LSP's baseline partition distribution is 
``hardwired'' into its pmf and is obtained when $p_\text{\,lsp}(\partition \mid \baselinepart, 
\shrinkitem=0)$.  Note that this partition distribution is a special case of
\citet{jensen2008bayesian}'s distribution with mass parameter fixed at $1$.

\section{Shrinkage Partition Distribution}
\label{sec:spd}

\subsection{Probability Mass Function}
\label{sec:spd_pmf}

We now present the pmf of our SP distribution, which incorporates key strengths 
of both the CPP and the LSP, addresses some of their limitations, and adds modeling flexibility. Like the CPP and the
LSP, the SP distribution uses an anchor partition $\baselinepart$.
Both the CPP and the LSP have a single, real-valued shrinkage
parameter $\shrinkitem \geq 0$, with larger shrinkage values producing higher concentrations of
probability mass for partitions that are close to $\baselinepart$. In contrast,
the SP distribution uses a vector-valued shrinkage $\shrink
= (\shrinkitem_1,\ldots,\shrinkitem_n)$ 
to permit item-specific knowledge and flexibility in how close realizations from the SP distribution
are to the anchor partition $\baselinepart$.  
That is, $\shrinkitem_i \ge 0$ is idiosyncratic to the $i^{\text{th}}$ item, although a modeler
could simplify $\shrink$'s structure by making each entry equal or choose to have
items in the particular subset of $\baselinepart$ all share the same shrinkage value (i.e., 
$\shrinkitem_i=\shrinkitem_j$ if $\baselineitem_i=\baselineitem_j$).

Like many other partition distributions, including the LSP, we adopted a sequential allocation 
construction for the SP distribution, with an
important caveat.  As it was originally defined, the LSP distribution implicitly uses a
fixed permutation $\perm = (1,\ldots,n)$, which makes the data analysis
dependent on the ordering of the data. In some contexts (e.g., time series),
the item order may be meaningful.  But most commonly, the permutation
$\perm$ would be viewed as a nuisance parameter.  Lacking prior knowledge
for $\perm$, we recommend using the uniform distribution on $\perm$, that is, $p(\perm)=\frac{1}{n!}$. Marginalizing over
$\perm$ has the effect of making the data analysis invariant to the order
of the data.
We also place a practical constraint on the baseline partition distribution $\baselinedist$: it  
must have an explicit allocation rule $\prb$ that conditions on previously allocated items.
Thus $\baselinedist(\partition) =
\prod_{k=1}^{n}  \prb(\partitionitem_{\permitem_{k}} \mid \partitionitem_{\permitem_{1}}, \ldots, \partitionitem_{\permitem_{k-1}} \,) =  
\prod_{k=1}^{n}  \prb(\partitionitem_{\permitem_{k}} \mid \partitionupto{k-1} \,)$,
where $\partitionupto{k-1} = \partitionitem_{\permitem_{1}}, \ldots, \partitionitem_{\permitem_{k-1}}$.
The function $\prb$ may well have other parameters (e.g., a concentration parameter)
but we suppress these here for generality and simplicity.
The function $\baselinedist$ may be an exchangeable partition probability function (EPPF) from, for example, 
the Ewens distribution \citep{ewens1972sampling, pitman1995exchangeable}, but we 
are not limited to exchangeable priors and only require a sequential allocation rule.  We call 
this sequential allocation rule $\prb$ a conditional allocation probability function (CAPF).

The CPP utilizes a distance function between $\partition$ and $\baselinepart$ 
such as the Binder 
loss \citep{binder.1978} or the variation of information 
\citep{meilua2007comparing,wade2018bayesian}. 
Analogously, the SP's anchor 
partition distribution uses a function inspired by the 
general form of Binder loss \citep{dahl2022search}.  This 
function, denoted $\text{Pr}_\text{a}$, rewards allocations which agree with 
$\baselinepart$ while simultaneously penalizing or rewarding larger clusters
through a real-valued grit parameter $\psi$.
The function is presented in CAPF form:
\begin{align} 
\label{Eq:SPAnchorParts}
\text{Pr}_\text{a}&(\partitionitem_{\permitem_k}=c \mid \partitionupto{k-1}, \baselinepart, \shrink, \perm, \psi ) 
\propto
& \exp \left( \frac{\shrinkitem_{\permitem_k}}{k-1} \sum_{j=1}^{k-1}  \shrinkitem_{\permitem_j} \, \text{I}\{ 
\partitionitem_{\permitem_j} = c \} \left( \text{I}\{ \baselineitem_{\permitem_j} = \baselineitem_{\permitem_k} \} - \psi \right) 
\right)
\end{align}
for $c \in \{ \partitionitem_{\permitem_1}, \ldots, \partitionitem_{\permitem_{k-1}}, \left| \{ \partitionitem_{\permitem_{1}}, \ldots, \partitionitem_{\permitem_{k-
1}} \} \right| +1 \}$.
Notationally, $\left| \{ \partitionitem_{\permitem_{1}}, 
\ldots,\partitionitem_{\permitem_{k-1}} \} \right|$ 
is the number of clusters before allocating the $\permitem_{k}^{\text{\ th}}$ item.
The pmf of the SP distribution is: 
\begin{equation} \label{Eq:SPDpmfSecond-old}
p_\text{sp}(\partition \mid \baselinepart, \shrink, \perm, \psi, \baselinedist)
= \prod_{k=2}^{n}
\text{Pr}_\text{sp}(\partitionitem_{\permitem_k} \mid \partitionupto{k-1}, \baselinepart, \shrink, \perm, \psi, \baselinedist ),
\end{equation}
where 
\begin{align} \label{eq:spCAPF}
    \text{Pr}_\text{sp}(\partitionitem_{\permitem_k} &\mid \partitionupto{k-1}, \baselinepart, \shrink, \perm, \psi, \baselinedist ) \propto
\text{Pr}_\text{b}(\partitionitem_{\permitem_k} \mid \partitionupto{k-1} )
\times
\text{Pr}_\text{a}(\partitionitem_{\permitem_k} \mid \partitionupto{k-1}, \baselinepart, \shrink, \perm, \psi ).
\end{align}
We note that, to simplify notation, we are conditioning on $\baselinepart$, $\shrink$, and $\perm$, 
however, the probabilities defined in (\ref{Eq:SPAnchorParts}) 
-- (\ref{eq:spCAPF}) are conditionally dependent on only some elements of these vectors.
For a random partition $\partition$, we use the notation 
$\partition \sim \text{SP}(\baselinepart,\shrink,\perm,\psi,\baselinedist)$ to denote the partition has a SP 
distribution with an anchor partition $\baselinepart$, a shrinkage vector $\shrink$, a permutation 
$\perm$, and a baseline distribution $\baselinedist$.  
When marginalizing over the permutation 
parameter with a uniform prior (thereby making the distribution invariant to the ordering of the data), we use the notation 
$\partition \sim \text{SP}(\baselinepart,\shrink,\psi,\baselinedist)$. Similar to the revised LSP introduced in Section \ref{sec:lsp}, we note that invariance to the ordering of the data does not imply that the SP is exchangeable.

From (\ref{Eq:SPDpmfSecond-old}), it is clear that the allocation of items to subsets 
in a partition are sequential.  However, the SP distribution is similar in form to the CPP in 
that 
the allocation of each item depends on the baseline partition distribution $\baselinedist$
and its compatibility with the anchor partition $\baselinepart$.
The CPP, however, applies this trade-off globally, whereas our SP distribution does so on a 
sequential, item-by-item basis.
At the $k^{\text{th}}$ step, note that there are only $\left| \{ 
\partitionitem_{\permitem_{1}}, \ldots, \partitionitem_{\permitem_{k-1}} \} \right| +1$ 
possible allocations
for $\sigma_k$.  Thus, the normalizing constant needed for (\ref{eq:spCAPF})
is readily computed for
any $n$, which easily permits posterior inference on any of the parameters
in the SP distribution using standard MCMC
techniques.

\newcommand{\gritloss}{%
Recall that the CPP allows for different distance functions and the chosen loss function implicitly favors some partitions over others.
For example, the Binder loss tends to produce many small clusters whereas VI tends to produce fewer clusters
\citep{rastellifriel2018,dahl2022search}.
Our grit parameter $\psi$ in (\ref{Eq:SPAnchorParts}) provides a similar mechanism to control the propensity for small clusters.
A key difference, however, is that our grit parameter $\psi$ is continuous and a prior can be placed on it, whereas the CPP's distance function must be chosen and fixed.}\gritloss{}

\subsection{Baseline Distributions} \label{sec:baselines}

The baseline distribution $\baselinedist$ is a key component of both the CPP and SP, which affords flexibility to the modeler.  Here we 
discuss a few obvious choices for $\baselinedist$ and emphasize that others may be desired for any given situation.
Conditional allocation probability functions are one way to characterize probability 
mass functions for partition distributions and are used in the SP distribution.  CAPFs rely on items being sequentially allocated 
to clusters in the partition.  Although most partition distributions can be viewed as 
sequentially allocated processes, those that cannot may not have tractable CAPFs.  
In this section, we examine a few partition distributions that could serve as a baseline 
distribution for the SP distribution and provide their associated CAPFs.

The two most common partition distributions are the Ewens 
\citep{ewens1972sampling, pitman1995exchangeable} 
and the Ewens-Pitman \citep{pitman1997two} distributions 
(also referred to as the one-parameter and two-parameter Chinese restaurant process,
respectively).
The CAPF for the Ewens-Pitman partition distribution is:
\begin{align} \label{EQ:Ewens-Pitman}
 \prb(\partitionitem_{\permitem_k} = c \ &| \ 
 \partitionupto{k-1},\alpha, \delta,\perm) &=
\begin{cases}
\frac{\left(\sum^{k-1}_{j=1} I(\partitionitem_{\permitem_{j}} = c)\right)-\delta}{k-1+\alpha} \hspace{2.3ex} \text{ 
for } c \in \{\partitionitem_{\permitem_1},\ldots,\partitionitem_{\permitem_{k-1}}\}  \\
\frac{\alpha + \delta \left| \{\partitionitem_{\permitem_1},\ldots,\partitionitem_{\permitem_{k-1}}\} 
\right|}{k-1+\alpha} \hspace{2ex} \text{ for } c = 
\left| \{\partitionitem_{\permitem_1},\ldots,\partitionitem_{\permitem_{k-1}}\} \right| +1
\end{cases}
\end{align}
Although this distribution is invariant to item allocation order, we include a permutation 
parameter in the CAPF for notational consistency.
The CAPF of the Ewens distribution can be obtained by setting $\delta=0$ in (\ref{EQ:Ewens-Pitman}).
Both the Ewens and Ewens-Pitman distributions have the ``rich-get-richer'' property which can be 
adjusted to some extent in the Ewens-Pitman distribution through the discount parameter $\delta$.
We use $\text{CRP}(\alpha)$ to denote the Ewens-Pitman distribution with concentration $\alpha$ and discount $\delta = 0$.

The uniform partition (UP) distribution is exchangeable and is a very simple partition distribution.
To be used as a baseline distribution for the SP distribution, its CAPF is needed.  To our knowledge,
the sequential allocation rule for the uniform partition distribution is not contained in the literature elsewhere
and we provide it here.
First, the total number of items to be partitioned, denoted $\numi$, must be fixed and known.  Then, at 
the $\permitem_k^{\text{\ th}}$ item's allocation, one must know the number of subsets 
allocated thus far in 
the partition (i.e., $| \{  \partitionitem_{\permitem_1},\ldots,\partitionitem_{\permitem_{k-
1}} \} |$) and also the number of items left to allocate (i.e., $n -k +1$).  Now consider an extension of the Bell number $B(a,b)$, defined recursively by the 
following formulas: $B(0,b)=1$ and $B(a+1,b)=b 
B(a,b)+B(a,b+1)$, for all $a$, $b \in \mathbb{N}$ (the nonnegative integers).  We note that 
$B(a) \equiv B(a,0)$ is the $a^{\text{th}}$ Bell number; additional information about the 
extension of the Bell number is included in Appendix \ref{sec:eBells}.
The CAPF for the UP distribution is:
\begin{align} \label{EQ:Uniform}
\prb(\partitionitem_{\permitem_k} = c \ | \ 
\partitionupto{k-1},n,\perm) =
\begin{cases}
\frac{B\left(n-k, \, | \{  \partitionitem_{\permitem_1},\ldots,\partitionitem_{\permitem_{k-1}} 
\} | \right)}{B\left(n-k+1, \, | \{ 
\partitionitem_{\permitem_1},\ldots,\partitionitem_{\permitem_{k-1}} \} |\right)} \hspace{2.4ex} \text{ for } 
c \in \{\partitionitem_{\permitem_1},\ldots,\partitionitem_{\permitem_{k-1}}\}  \\
\frac{B\left(n-k, \, | \{  \partitionitem_{\permitem_1},\ldots,\partitionitem_{\permitem_{k-1}} 
\} | +1 \right)}{B\left(n-k+1, \, | \{ 
\partitionitem_{\permitem_1},\ldots,\partitionitem_{\permitem_{k-1}} \} |\right)} \hspace{2.4ex} \text{ for } 
c = \left| \{\partitionitem_{\permitem_1},\ldots,\partitionitem_{\permitem_{k-1}}\} \right| +1,
\end{cases}
\end{align}
for $k \in \{2,\ldots,n\}$.  When $k=1$, the CAPF for the UP distribution simply equals 1.

The Jensen-Liu partition (JLP) distribution  was introduced by \cite{jensen2008bayesian} 
and 
later named in \cite{casella2014cluster}.  It places a uniform probability of allocating to any 
existing cluster and a distinct  probability of forming a new cluster. 
Unlike the previously mentioned
distributions in this section, the JLP is not exchangeable.
The CAPF for the JLP (endowed with a permutation parameter) is:
\begin{align} \label{EQ:Jensen-Liu}
\prb(\partitionitem_{\permitem_i} = c \ | \ 
\partitionupto{i-1},\alpha,\perm) = &
\begin{cases}
\frac{1}{| \{ \partitionitem_{\permitem_1},\ldots,\partitionitem_{\permitem_{i-1}} \} |+\alpha} 
\hspace{2.4ex} \text{ for } c \in \{\partitionitem_{\permitem_1},\ldots,\partitionitem_{\permitem_{i-1}}\}  \\
\frac{\alpha}{| \{ \partitionitem_{\permitem_1},\ldots,\partitionitem_{\permitem_{i-1}} \} 
|+\alpha} \hspace{2.4ex} \text{ for } c = 
\left| \{\partitionitem_{\permitem_1},\ldots,\partitionitem_{\permitem_{i-1}}\} \right| +1.
\end{cases}
\end{align}

There are other possible baseline partition distributions.
\cite{dahl2017random} introduces the Ewens-Pitman attraction (EPA) distribution, which 
generalizes the Ewens-Pitman distribution when pairwise distances (or similarities) between items are known.  
Using the EPA 
distribution would allow one to incorporate covariates into the partitioning of items by 
converting the covariates to similarities between items.
Another recent addition to partition distributions is the Allelic partition distribution 
\citep{betancourt2022prior}. This distribution was developed specifically to have the 
microclustering property (that is, partitions contain many clusters in which there are relatively 
few items in each).  This partition behavior is in stark contrast to the Ewens distribution, which 
favors large cluster sizes.
Product partition models (\citealp[PPMs][]{hartigan1990partition}) can be used as a baseline in the SP distribution.
Suitable extensions of 
PPMs incorporate covariates 
\citep{muller2011product,park2010bayesian} or those that add spatial structure with covariates \citep{page2016spatial}.

\section{Effect of Shrinkage and Grit Parameters}
\label{sec:prior_exploration}

An intuitive way to think about the shrinkage partition (SP) distribution
is to view it as a compromise between a baseline partition 
distribution $\baselinedist$ and the anchor partition $\baselinepart$.
This compromise is controlled by the shrinkage parameter $\shrink$ and the grit parameter $\psi$.
In this section, we investigate in more detail how these parameters influence the SP distribution.
Proofs of the theorems are found in Appendix \ref{apx:blc}.

\subsection{Extremes and Smooth Evolution Between Extremes}
\label{sec:anchor_consistency}

The degree of compromise between the baseline partition 
distribution $\baselinedist$ and the anchor partition $\baselinepart$ is governed by the shrinkage parameter $\shrink = (\shrinkitem_{1}, \ldots, \shrinkitem_{n})$.
In one extreme case, where each idiosyncratic shrinkage $\shrinkitem_{i}$ is 
relatively large, the SP distribution assigns probability mass to partitions that are very 
similar to the anchor partition $\baselinepart$ and, in the limit, the distribution
becomes a point mass distribution at the anchor partition 
$\baselinepart$.
In the other extreme case, i.e., when the idiosyncratic shrinkages $\shrinkitem_{i}$ all equal zero, 
the SP reduces to the baseline distribution $\baselinedist$ and the
anchor partition $\baselinepart$ has no influence in the SP's pmf.
We formally present these properties below.

\newcommand{\PropOne}{
Let $\bm{\rho}_1$ be the partition with all $n$ items assigned to a single cluster,  $\bm{\rho}_n$ be the partition with $n$ items each assigned to a unique cluster,  $\partition \sim \text{SP}(\baselinepart, \shrink, \psi, \baselinedist)$ with $\baselinedist(\baselinepart) > 0$, and 
$\shrink = \shrinkitem \times (1,\ldots,1)$:
\begin{align*}
    &\text{a)} \ \text{if} \  \shrinkitem=0, \text{ then } \ \partition \sim \baselinedist. 
    & \text{[Diffusion to the baseline dist'n]} \\
    &\text{b)} \ \text{if } \psi \in (0,1) \text{ then } \lim_{\shrinkitem \to \infty} \text{Pr} \,(\partition=\baselinepart) = 1 
    & \text{[Anchor partition consistency]}\\
    &\text{c)} \ \text{if } \psi \in (-\infty,0) \text{ then } \lim_{\shrinkitem \to \infty} \text{Pr} \,(\partition=\bm{\rho}_1) = 1 \\
    &\text{d)} \ \text{if } \psi \in (1,\infty) \text{ then } \lim_{\shrinkitem \to \infty} \text{Pr} \,(\partition=\bm{\rho}_n) = 1. 
\end{align*}
}
\begin{theorem}
\label{thm:BLC}
\PropOne
\end{theorem} 
\noindent
Recall that $\partition$ and $\baselinepart$ are vectors of cluster labels.
When statements such as $\partition=\baselinepart$ are made, we imply that they are
in the same equivalence class, that is, their cluster
labels encode the same partition.

\newcommand{\desirableproperty}{Note that as the shrinkage $\shrinkitem$ goes to infinity, the desirable behavior that $\text{Pr} \,(\partition=\baselinepart) = 1$ only occurs when $\psi \in (0,1)$.
Of course, for finite $\shrinkitem$, there could be situations in
which better model fit is obtained by  relaxing the restriction that $\psi \in
(0,1)$. We feel, however, that it would be better in such situations to revisit
the choice for the baseline distribution $p_b$ to influence the number and size of clusters.
As such, we recommend restricting $\psi$ to the unit interval and do so throughout the rest of the paper.}\desirableproperty{}

Between the extreme cases for the shrinkage $\shrinkitem$ (i.e., zero and infinity) and under certain conditions, the probability of the anchor partition
is monotonically increasing in $\shrinkitem$.
\newcommand{\theoremOne}{For any $\delta > 0$ and $\baselinedist$ such that $0 < \baselinedist(\baselinepart) < 1$, if $\partition_1 \sim \text{SP}(\baselinepart, \shrink, \psi, \baselinedist)$ 
and $\partition_2 \sim \text{SP}(\baselinepart, \shrink + \delta, \psi, \baselinedist)$ with 
$\shrink = \shrinkitem \times (1, \dots, 1)$, $0 \leq \shrinkitem<\infty$ and $\psi \in (0,1)$, 
then $\text{Pr}\,(\partition_1 = \baselinepart) < \text{Pr}\,(\partition_2 = \baselinepart)$.}
\begin{theorem}
\label{thm:monotonic_pm}
\theoremOne
\end{theorem}
\noindent
This theorem implies that $\shrinkitem$ controls how ``close'' in probability the SP 
distribution is to the anchor partition $\baselinepart$.  Additionally, this closeness to 
$\baselinepart$ is strictly increasing as $\shrinkitem$ increases.  

Whereas Theorem \ref{thm:monotonic_pm} illustrates monotonically increasing probability
for the anchor $\baselinepart$ as a function of increasing $\shrinkitem$,
in fact the entire distribution is getting closer to a point mass distribution at $\baselinepart$.
In the following theorem we show the Kullback-Leibler (KL) divergence and the total variation (TV) distance between the SP 
distribution and a point mass distribution at the anchor decreases monotonically with increasing 
$\shrinkitem$.  We also show that 
the expectation of the Rand index (RI), a measure of agreement between partitions, converges to 1 as $\shrinkitem$ goes to infinity, implying the partitions are identical asymptotically.
\newcommand{\theoremTwo}{Consider two partition distributions: i) $p_{\baselinepart}$, 
a point mass at the anchor 
partition $\baselinepart$, and ii) $\text{SP}(\baselinepart, \shrink, \psi, \baselinedist)$ 
such that
$\shrink = \shrinkitem \times (1, \dots, 1)$ 
with $0 \leq \shrinkitem < \infty$, $\psi \in (0,1)$, and $0 < \baselinedist(\baselinepart) < 1$.  If $\baselinepart \sim p_{\baselinepart}$ and $\partition \sim \text{SP}(\baselinepart, \shrink, \psi, \baselinedist)$ then:
\begin{enumerate}[a)]
\item $D_{KL}\big(\baselinepart, \partition \big)$ is strictly decreasing as $\shrinkitem$ increases.
\item 
$D_{TV}\big(\baselinepart, \partition\big)$
is strictly decreasing as $\shrinkitem$ increases.
\item $E_{\baselinepart,\partition}\left[RI\big(\baselinepart, \partition\big)\right] \to 1$ as $\shrinkitem \to \infty$.
\end{enumerate}
}
\begin{theorem}
\label{thm:kullback}
\theoremTwo
\end{theorem}
\noindent Although not a true metric, the Kullback-Leibler divergence provides a sense of how 
relatively ``close'' two distributions are to each other.  This theorem reinforces the concept 
that the SP distribution is a compromise between the baseline distribution $p_{\text{b}}$ and 
the anchor partition $\baselinepart$.  When $\shrinkitem=0$, the SP distribution is identical to 
$p_{\text{b}}$, and 
$D\big(\baselinepart, \partition \big)$ is at its 
maximum.  Then, as $\shrinkitem$ increases, 
$D\big(\baselinepart, \partition \big)$ 
strictly decreases, or in other words, the SP distribution moves away from 
$p_{\text{b}}$ and toward the point mass distribution at $\baselinepart$.

\begin{figure}[tb]
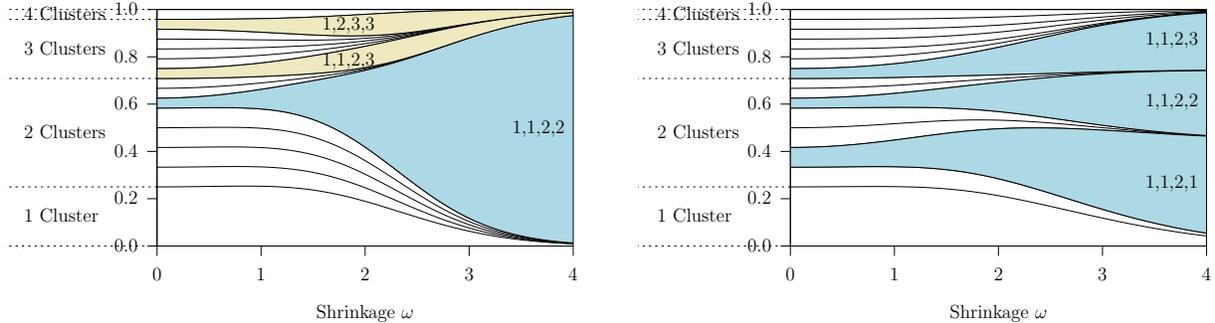

     \centering
     \begin{subfigure}[b]{0.49\textwidth}
         \centering
         \resizebox{0.95\linewidth}{!}{\input{img/stackplot_1}}
         \label{fig:LocScaleConst1}
     \end{subfigure}
     \hfill
     \begin{subfigure}[b]{0.49\textwidth}
         \centering
         \resizebox{0.95\linewidth}{!}{\input{img/stackplot_2}}
         \label{fig:LocScaleConst2}
     \end{subfigure}
     \caption{Left: The probabilities of all possible partitions for $n=4$ items evolving as the shrinkage $\shrinkitem$ increases.  As discussed in Section \ref{sec:anchor_consistency}, the areas highlighted in blue show that the probability of the anchor partition begins to dominate as $\shrinkitem$ increases.
     Right: The evolution of SP's probabilities when $\shrinkitem_{4}=0$ 
     and the others increase.  As discussed in Section \ref{sec:generalizations}, the blue areas highlight the limiting partitions.}
  \label{fig:LocScaleConst}
\end{figure}

Theorems \ref{thm:BLC}, \ref{thm:monotonic_pm}, and \ref{thm:kullback} provide intuition for the shrinkage parameter,
which is reinforced with an illustrative example.
Consider the SP distribution in which we marginalize over a uniform prior on the permutation $\perm$, we fix the grit $\psi$ at 0.3, $\baselinedist$ is the CRP(1),
and $\shrink = \shrinkitem \times (1,\ldots,1)$.
The left plot in Figure \ref{fig:LocScaleConst} shows the evolution of the SP's probabilities of an anchor partition as a function of increasing $\shrinkitem$.
The plot shows the probabilities for the $B(4) = 15$ partitions
of $n=4$ items that are stacked so that, if a vertical line segment were drawn between 0 and 1 at some value 
of $\shrinkitem$, then the line segment would be divided into 15 pieces, with the length of each piece representing 
the probability of one of the $15$ partitions.  The length of the vertical line segment in the blue 
region denotes the probability of the anchor partition $\baselinepart = (1,1,2,2)$
which, when $\shrinkitem = 0$, is determined solely from the baseline distribution $p_{\text{b}}$
and then converges to $1$ as $\shrinkitem$ goes to infinity.
The plot visually depicts an example of anchor partition consistency, diffusion to the baseline distribution, and the smooth evolution between these two extremes.
Also note that the partitions $(1,1,2,3)$ and $(1,2,3,3)$ are shaded in yellow.  These partitions are similar to the $\baselinepart = (1,1,2,2)$
and therefore have increasing probability for small shrinkage $\shrinkitem$ which eventually gives way to the anchor $\baselinepart = (1,1,2,2)$ for large $\shrinkitem$.
The upshot of these theorems and example is that the shrinkage parameter
$\shrink$ smoothly controls the compromise between a baseline partition 
distribution $\baselinedist$ and the anchor partition $\baselinepart$.
The SP distribution inherits clustering properties
of $\baselinedist$ which are tempered by the anchor
partition $\baselinepart$ based on the values of $\shrink$. 

\newcommand{\muelleretal}{This compromise behavior is reminiscent of
\citet{mueller2004} who showed how to define a novel distribution function as a
convex linear combination of two distributions, e.g., $(1-\epsilon)F + \epsilon
F_0$, where $\epsilon \in [0,1]$ controls the degree of compromise between $F$
and $F_0$. In our SP distribution, the compromise parameter is our vector-valued
shrinkage parameter $\shrink$, and it induces a total variation distance that
takes values in $[0,1]$. As shown in the proof of Theorem \ref{thm:kullback},
the total variation distance $D_{TV}$ between the SP distribution and the
point mass distribution at the anchor $ \baselinepart$ equals one minus the
probability of the anchor $\baselinepart$, i.e., in the left plot of Figure \ref{fig:LocScaleConst}, one minus the length of a vertical line
segment in blue for any given shrinkage $\shrinkitem$. So, in this example with
$n=4$, $D_{TV}$ is $0.96$ when $ \shrinkitem = 0$, decreases to $0.68$ when $
\shrinkitem = 2$, and equals $0.04$ when $\shrinkitem = 4$. In a situation with
$n$ being much larger, the total variation distance is still easily computed
for a fixed permutation and readily approximated through Monte Carlo methods for
a random permutation.}\muelleretal{}

\subsection{Idiosyncratic Shrinkage Parameters}
\label{sec:generalizations}

Recall that the SP distribution is parameterized with a vector-valued shrinkage
parameter $\shrink = (\shrinkitem_1, \ldots, \shrinkitem_n)$. Whereas the CPP
and LSP have only a univariate shrinkage parameter $\shrinkitem$ to inform
the compromise between the baseline distribution $p_{\text{b}}$ and the anchor
partition $\baselinepart$, our SP distribution allows for idiosyncratic
shrinkages. The allowance of item-specific shrinkages $\shrinkitem_{1},\ldots,\shrinkitem_{n}$ offers
unique flexibility in  prior elicitation that is consequently not possible in
existing distributions. Specifically, consider a data analysis scenario where
there is strong prior knowledge regarding  the clustering of a few items, but
little or no prior knowledge regarding the clustering  of the remaining items.
In this case, the SP prior may avoid inadvertently imposing prior influence on
less-understood  relationships by setting corresponding shrinkage values to be
relatively small or, at the extreme, $0$ itself. Conversely, strong \textit{a
priori} clustering information about a few items can be expressed  without fear
of unduly affecting items which are not well understood.

For illustration, consider again an SP distribution in which $n=4$, $\perm$
is integrated  out of the model, $\psi=0.3$, and $\baselinedist$ is $\text{CRP}(1)$.
Recall the scenario in Section
\ref{sec:anchor_consistency} with the anchor $\baselinepart = (1, 1, 2, 2)$ and
$\shrink = \shrinkitem \times (1, 1, 1, 1)$.  Now assume instead there is
no prior knowledge about the clustering of the fourth item (but we do have an
indication that the first and second items are clustered together and are not
clustered with the third item).  That is, we are indifferent to $\baselinepart
= (1, 1, 2, 1)$, $\baselinepart = (1, 1, 2, 2)$, and $\baselinepart = (1,
1, 2, 3)$, apart from beliefs about the number and size of clusters encoded within the baseline distribution $p_{\text{b}}$ and the grit parameter $\psi$.
In this case, we let $\shrink = \shrinkitem \times (1, 1, 1,
0)$ and any of these three values for the anchor $\baselinepart$ yield exactly the same
partition distribution. The evolution of probabilities as $\shrink$ increases is
illustrated in the right plot of Figure \ref{fig:LocScaleConst}.  Notice that
$(1, 1, 2, 1)$, $(1, 1, 2, 2)$, and $(1, 1, 2, 3)$ all retain probability as $\shrinkitem$ goes to infinity.

This example displays another key feature of the SP distribution when $\shrink$ contains 
both zero and non-zero values.
We highlight that the SP distribution is invariant to the \textit{a priori} 
clustering of items with shrinkage parameter set to 0, i.e., $\shrinkitem_{i} = 0$.
\newcommand{\indifference}{For example, since $\shrinkitem_4 = 0$, the probabilities for the SP distribution specified in the right pane of Figure 
\ref{fig:LocScaleConst} would remain unchanged if $\baselinepart = (1,1,2,1)$ or $\baselinepart = (1,1,2,3)$ instead of 
$\baselinepart = (1,1,2,2)$.}\indifference{} 
\newcommand{\theoremThree}{If $\partition_1 \sim \text{SP}(\baselinepart,\shrink,\perm,\psi,\baselinedist)$ 
and $\partition_2 \sim \text{SP}(\baselinepart^*,\shrink,\perm,\psi,\baselinedist)$, where $\baselinepart$ 
and $\baselinepart^*$ are anchor partitions such that for every $i$ and $j$ where $\shrinkitem_i > 0$ and $\shrinkitem_j > 0$, $\baselineitem_i=\baselineitem_j$ if and only if $\baselineitem_i^*=\baselineitem_j^*$,
then $\partition_1$ and $\partition_2$ are equal in distribution.}
\begin{theorem}
\label{thm:zeroweights}
\theoremThree
\end{theorem}
\noindent The proof of Theorem \ref{thm:zeroweights} in Appendix \ref{apx:blc}
essentially says that an item with a shrinkage parameter of zero 
can be placed into any cluster in the anchor partition without affecting the probabilities of 
the SP distribution.  This fits nicely with an intuitive Bayesian 
interpretation of having a shrinkage parameter set to zero and relieves the modeler from making an arbitrary choice that could affect the analysis.
The behavior of the SP distribution is 
demonstrated in the right plot in Figure \ref{fig:LocScaleConst}, and gives rise to what we refer to as ``limiting 
partitions.''
The definition of a limiting partition is:
\begin{definition} \label{def:limiting}
Let $\partition \sim \text{SP}(\baselinepart, \shrink, \psi, \baselinedist)$ with
$\shrink = \shrinkitem \times (s_{1}, \dots, s_{n})$ for scalars 
$s_{i} \in [0,\infty)$.
A partition $\partition$ which has non-zero probability as $\shrinkitem \to \infty$ 
is a limiting partition.
\end{definition}
\noindent When one or more idiosyncratic shrinkage parameters are set to zero and assuming $\baselinedist(\baselinepart)>0$, the SP 
distribution does not reduce to a point mass at $\baselinepart$ as $\shrinkitem$ becomes large.
However, $\baselinepart$ remains a limiting partition and, if at least two items have $s_{i}>0$, 
then the number of limiting partitions is less than the 
number that are possible under $\baselinedist$.  
In the limit, this allows the SP distribution to force some 
items to be clustered and other items to be separated, and 
leave the remaining items to be randomly clustered according to $\baselinedist$.
The following theorem characterizes the limiting partitions. 
\newcommand{\propTwo}{Consider any baseline distribution $\baselinedist$, anchor partition $\baselinepart$ and fixed partition  
$\bm{\rho}$, such that $\baselinedist(\baselinepart) > 0$ and $\baselinedist(\bm{\rho}) > 0$.  For any 
$\shrink = \shrinkitem \times (s_{1}, \dots, s_{n})$ with $s_{i}$ either 0 or 1, and $\psi \in (0,1)$, define 
$\mathcal{Q}$ to be the set of index pairs $(i,j)$ such that $\omega_i > 0$, $\omega_j > 0$, and 
$\baselineitem_i = \baselineitem_j$.  If $\rho_i = \rho_j$ for all index pairs in $\mathcal{Q}$, then 
$\bm{\rho}$ is a limiting partition of $\partition \sim \text{SP}(\baselinepart, \shrink, \psi, \baselinedist)$.}
\begin{theorem}
    \label{prop:limiting}
\propTwo
\end{theorem}

One interesting quantity of the SP distribution is the number of limiting partitions which 
exist.  
We know the number of possible partitions for $n$ items with no constraints is the 
$n^{\text{th}}$ Bell 
number, and on the other extreme, there is only one partition (a point mass) when all values in 
$\shrink$ go to infinity.  
However, when only some elements in $\shrink$ go to infinity, the number of limiting partitions is 
between those two extremes.  Theorem \ref{prop:NumLimParts}
gives the number of limiting partitions. 

\newcommand{\propThree}{Let $\partition \sim \text{SP}(\baselinepart,\shrink,\perm,\psi,\baselinedist)$, with 
$\baselinedist(\partition) > 0$ for all possible $\partition$ and 
$\shrink = \shrinkitem \times (s_{1},\ldots,s_{n})$, where 
$s_{i}$ is either 0 or 1 and $\psi \in (0,1)$.  Let 
$a=n-\sum_{i=1}^{n} s_{i}$
be the number of items with shrinkage equal to zero.
Let $b$ be the number of clusters in $\baselinepart$ having at least one item with $s = 1$.
As $\shrinkitem \to \infty$, the number of limiting partitions in the SP 
distribution is $B(a,b)$}
\begin{theorem} \label{prop:NumLimParts}
\propThree---an extension of the Bell numbers from Section \ref{sec:baselines}. 
\end{theorem}

\subsection{Prior Elicitation for Shrinkage and Grit Parameters}
\label{sec:simulateprior}

\newcommand{\simulateprior}{The previous subsections investigated the effects as elements of the
shrinkage vector $\shrink$ go to infinity and, when the number of items $n$
permits exhaustive enumeration, the effects for finite values of $\shrink$.
More generally, the effect of a particular shrinkage value (e.g., $\shrinkitem = 4$)
will depend on other factors, such as the sample size $n$, the anchor partition $\bm{\mu}$, and the grit $\psi$.
When enumeration is not feasible, we recommend simulation when
eliciting priors on (or fixed values for) the shrinkage and grit parameters.
We suggest specifying a two-dimensional grid of values for the shrinkage and grit parameters
and then sampling from the SP distribution under each combination of values.
We then recommend computing, for example, Monte Carlo estimates of the total variation distance, the
expectation of the Rand index, and the expectation of the cluster entropy
--- defined as the negation of the sum of the relative cluster sizes times the
natural logarithm of the relative cluster sizes --- for each combination of shrinkage and grit values.
The modeler can pick priors for shrinkage and grit values that are compatible with
\textit{a priori} beliefs for these quantities.
We demonstrate these ideas in Appendix \ref{sec:clusterestimation}.}\simulateprior{}

\section{Dependent Random Partitions}
\label{sec:dependRPs}

\subsection{Hierarchically-Dependent Random Partitions}
\label{sec:hierarchical_model}

Over the past several decades, Bayesian hierarchical models 
have been used extensively with a great deal of 
success.  Dependence among datasets can be induced between similar populations by placing a
common prior distribution on parameters of interest, e.g., means and variances, and then placing a hyperprior distribution on
the hyperparameters of the common prior distribution.  Using our SP distribution, these same principles can be 
applied to dependent partitions.
Specifically, consider $T$ separate collections of 
data, each with the same items to be clustered, and let
$\bm{\pi}_1$, \ldots, $\bm{\pi}_T$ be the latent partitions.
We can ``borrow strength'' in the estimation of these partitions
through the following hierarchical model:
\begin{equation}
\label{eq:hierarch}
\bm{\pi}_t \mid \baselinepart \sim \text{SP}(\baselinepart, \bm{\omega}, \perm_t, \psi, \baselinedist), \hspace{10ex} \baselinepart \sim p(\baselinepart).
\end{equation}
Any choice could be made for $p(\baselinepart)$,
e.g., a CRP or an SP with a ``best guess'' anchor.
There are likewise many choices for $\baselinedist$.
MCMC techniques can be used for posterior inference on hyperparameters because
the normalizing constant of the SP distribution is available.

The default choice for the 
shrinkage parameter might be $\shrink = \shrinkitem \times 
(1,\ldots,1)$ with a prior on the scalar $\shrinkitem$.
A more sophisticated prior could easily be adapted, 
such as, $\shrink = \shrinkitem \cdot 
s_t$, with $s_i \geq 0$.  Here $s_i$ would 
permit each partition to have varying degrees 
of dependence among the collection of the $T$ partitions.
The point is that the degree of dependence among $\bm{\pi}_1$, \ldots, $\bm{\pi}_T$
is governed by the shrinkage parameter $\shrinkitem$.  As with typical Bayesian hierarchical models, the model in (\ref{eq:hierarch})
allows the extremes of independence and exact equality among
the partitions $\bm{\pi}_1$, \ldots, $\bm{\pi}_T$
by setting $\shrinkitem = 0$ or $\shrinkitem \rightarrow \infty$, respectively.

The hierarchy shown in (\ref{eq:hierarch}) using the SP distribution
can be expanded upon.
For example, while analyzing neuroimaging data, 
\cite{song2023clustering} generalizes our approach.
In particular, they use an early version of our SP 
distribution in an innovative model to simultaneously 
cluster in two domains, namely, bi-clustering subjects
and conditions.  
\newcommand{\multiview}{Additionally, an emerging concept in the partition literature is that of \textit{multi-view} data \citep{duan2020latent, franzolini2023conditional, dombowsky2023product}.  Our hierarchical framework naturally allows for different data types (perhaps some low-dimensional and others high-dimensional) to share information using partitions with a common latent partition.}\multiview{}

\subsection{Temporally-Dependent Random Partitions}
\label{sec:temporal_model}

Modeling dependent partitions over time is also 
very natural using our SP distribution.
Again, assume we have partitions $\bm{\pi}_1$, \ldots, $\bm{\pi}_T$ for
the same items over time.
Recently \cite{page2022dependent}
presented a model to induce temporal dependence in partitions.  They argue that dependency 
should be placed directly on the partitions, as opposed 
to the approach in much of the literature which attempts to induce 
dependence on random measures.  We follow \citet{page2022dependent}'s philosophy and model dependence directly on the partitions 
using our SP distribution.  A basic time-dependent model is:
\begin{align} \label{eq:temporal}
\bm{\pi}_t \mid \bm{\pi}_{t-1}  \sim \text{SP}(\bm{\pi}_{t-1}, \bm{\omega}, \perm_t, \psi, p_b), \hspace{10ex} \bm{\pi}_1 & \sim p(\bm{\pi}_1),
\end{align}
for $t=2,\ldots,T$.  When data are evenly 
spaced in time, we envision setting 
$\shrink = \shrinkitem \times (1,\ldots,1)$ with 
$\shrinkitem$ a common parameter over time.  
However, unlike \cite{page2022dependent}, our framework is not restricted to data analysis involving equally-spaced time points. 
For example, the model is easily modified by parameterizing the shrinkage to be time-dependent, e.g., 
$\shrink_t = \frac{\shrinkitem}{d_t} \times (1,\ldots,1)$, where $d_t$ is the difference in time 
between time $t$ and time $t-1$.

\section{Empirical Application: Return to Education}
\label{sec:AppRTE}

In this section, we illustrate various uses of our
SP distribution, demonstrate the dependent partition models from the previous section, and compare results from related partition distributions.
Consider a regression model studying the relationship between earnings and education attainment.
The data were obtained from IPUMS CPS
\citep{ipums}.
In March of each year, a cross-section of individuals from the 50 United States and
the District of Columbia (D.C.) are surveyed regarding their employment and demographic 
information.  Data with
harmonized variables are available from the years 1994 to 2020.  Applying sensible filters for 
our application (e.g., only including working individuals)
yields 139,555 observations.  For simplicity, sampling weights are ignored for this 
demonstration.

Let $\bm{y}_{it}$ be the vector of the natural logarithm of the average hourly earnings for 
individuals in state $i=1, \ldots, n$ ($n=51$) in year $t=1,\ldots,T$ (with $t=1$ corresponding 
to the year 1994 and $T=27$).
Let $\bm{X}_{it}$ be a matrix of covariates containing a column of ones for the intercept and 
the following mutually-exclusive dummy
variables related to educational attainment: high school graduate, some
college, or at least a bachelors degree.  The matrix $\bm{Z}_{it}$ consists of non-education 
related covariates:
age, age$^2$, hours worked
weekly (\mbox{weeklyhrs}), weeklyhrs$^2$, and the following dummy variables: male, white,
Hispanic, in a union, and married. We consider the regression model 
$\bm{y}_{it} = \bm{X}_{it} \bm{\beta}_{it} + \bm{Z}_{it} \bm{\gamma}_{t} + \bm{\epsilon}_{it}$,
where the elements of $\bm{\epsilon}_{it}$ are independent and identically distributed normal 
with mean zero and precision $\tau_t$.
Conditional independence is assumed across states and years.  Note that $\bm{\gamma}_t$ and 
$\tau_{t}$ lack a subscript $i$ and are therefore common across
all states.  For a quick reference to the notation, we refer readers to Table \ref{tab:notation}
in Appendix \ref{sec:MCMC_Algo}.

Interest lies especially in estimating the state-year specific regression coefficients 
$\bm{\beta}_{it}$ regarding educational attainment.
Some state-year combinations have sufficient data (e.g., California has at least 497 
observations per year),
but some have limited data (e.g., D.C.\ has a year with only 20 observations) such that 
estimating $4 + 9 + 1 = 14$ parameters
would be very imprecise.  One solution is to combine data from small states, although such combinations 
may be \textit{ad hoc}.  Instead, in a given year $t$, we obtain
parsimony and flexibility by postulating
that the states are clustered by ties among $\bm{\beta}_{it}$ for $i=1,\ldots,n$.
Let $\partition_t = (\partitionitem_{1t},\ldots,\partitionitem_{nt})$ be the clustering 
induced by these ties,
where $\partitionitem_{it} = \partitionitem_{jt}$ if and only if 
$\bm{\beta}_{it} = \bm{\beta}_{jt}$.
For notational convenience, we label the first item $1$, use consecutive integers $1,\ldots,q_t$ 
for the
unique cluster labels in $\partition_t$, and let $\bm{\beta}^*_{1t},\ldots,\bm{\beta}^*_{q_tt}$ be 
the unique values for the regression coefficients.
Let $\bm{X}_{ct}$ be the
matrices obtained by vertically stacking the matrices $\bm{X}_{it}$ for all $i$ such that 
$\partitionitem_{it} = c$.
Likewise, $\bm{Z}_{ct}$ is obtained by vertically stacking
$\bm{Z}_{it}$ when $\partitionitem_{it} = c$
and let $\bm{y}_{ct}$ be the vector
obtained by concatenating $\bm{y}_{it}$ for all $i$ such that $\partitionitem_{it} = c$.

The joint sampling model for data $\bm{y}_{1t},\ldots,\bm{y}_{nt}$ is:
\begin{equation} \label{eq:like1}
p(\bm{y}_{1t},\ldots,\bm{y}_{nt} \mid \bm{\pi}_t, \bm{\beta}^*_{1t}, \ldots, \bm{\beta}^*_{q_tt}, 
\bm{\gamma}_t, \tau_t)
= \prod_{c=1}^{q_t} p(\bm{y}_{ct} \mid \bm{\beta}^*_{ct}, \bm{\gamma}_t, \tau_t),
\end{equation}
\begin{equation} \label{eq:like2}
\text{with\ \ } \bm{y}_{ct} \mid \bm{\beta}^*_{ct}, \bm{\gamma}_t, \tau_t \ \sim \ 
\text{N}\left( \bm{X}_{ct} \bm{\beta}^*_{ct} + \bm{Z}_{ct} \bm{\gamma}_t, \,  \tau_t \bm{I} \right),
\end{equation}
where $\bm{I}$ is an identity matrix and $\text{N}(\bm{a},\bm{B})$ represents a 
multivariate normal distribution with
mean $\bm{a}$ and precision matrix $\bm{B}$.

We use the following joint prior distribution for the other parameters as the product of
independent distributions.  T\newcommand{\precisionsareiid}{he precisions $\tau_1,\ldots,\tau_T$ in the sampling distribution are \textit{iid}
gamma with shape}\precisionsareiid{} $a_\tau = 1/0.361^2$ and rate $b_\tau = 1$, making the prior expected standard deviation
equal a preliminary guess of 0.361 obtained by exploratory data analysis.
The regression coefficient vector $\bm{\gamma}_t$
has a normal prior distribution with mean
$\bm{\mu}_\gamma = \bm{0}$ and precision $\bm{\Lambda}_\gamma = \bm{I}$.  Based
on preliminary explorations using ordinary least squares,
the priors for 
the cluster-specific regression coefficients
$\bm{\beta}^*_{1t},\ldots,\bm{\beta}^*_{q_t t}$ are independent and identically 
distributed normal with mean
$\bm{\mu}_\beta = (1.46, 0.15, 0.24, 0.41)^\prime$ and precision $\bm{\Lambda}_\beta$ is $100 \bm{I}$.

As the pmf of the SP distribution is
available in closed form, the full suite of standard
MCMC algorithms are available to sample from the posterior distribution. Our
approach for updating the cluster labels is a P\'olya urn Gibbs sampler based on
\citet{neal2000markov}'s Algorithm 8. One caveat is that the SP distribution is
not exchangeable, so its full pmf --- or, at least, the part
involving the current item and any item allocated after it according to
the permutation $\perm$ ---  must be computed.  Specific MCMC details are given in Appendix \ref{sec:MCMC_Algo}.

Recall that we have $T$ years with partitions $\partition_1, \ldots, \partition_T$.
We now discuss three specifications for the joint prior distribution for 
$\partition_1, \ldots, \partition_T$.
We first consider independent random partitions in Section \ref{sec:single} and use this model 
as a means to compare the
SP distribution with the CPP and the
LSP.  We then consider hierarchically and temporally dependent 
random partitions in
Sections \ref{sec:hierarchical} and \ref{sec:temporal}, respectively.

\subsection{Application Using Independent Partitions}
\label{sec:single}

In this section, we consider various prior distributions for 
$\partition_1, \ldots, \partition_T$ of the 
form $p(\partition_1, \ldots,\allowbreak\partition_T) = \prod_{t=1}^T p(\partition_t)$.
One's best guess for an unknown partition $\partition_t$ may be the partition given by the four US 
Census Regions, shown in Figure \ref{fig:regions} of Appendix \ref{sec:censusregions}.
Indeed, at one extreme, one may wish to fix $\partition_t$ to be these regions, effectively using a 
point-mass distribution $p(\partition_t)$ on the regions.
At the other extreme, one may disregard the regions and specify, for example, that 
$\partition_t \sim \text{CRP}(1)$ or $\partition_t \sim \text{JLP}(1)$.

We entertain compromises between these extremes.  In this subsection, we fix $\bm{\mu}$ to be the anchor
partition given by the U.S. Census Bureau regions and consider several
partition distributions informed by this ``best guess'' partition $\bm{\mu}$.
Specifically, we consider the shrinkage partition (SP) distribution,
the centered partition process (CPP), and the location-scale partition (LSP) distribution. 
Recall that the normalizing constant
is not known for the CPP and hence the shrinkage value $\omega$ must be fixed when
using standard MCMC methods.
In this subsection, for the sake of comparison with the CPP,  
we also fix the shrinkage for the SP and LSP.  
The specifics of how $\omega$ influences each distributional family differ
and it is not clear how to calibrate an $\omega$ for each distributional family
such that the influence of $\bm{\mu}$ is commensurate. Instead, we consider a
distribution-specific grid of values for $\omega$ and 
only present results for the best-fitting model of each distributional family.
We show results for models using the following prior partition distributions: i. ``Location-Scale Partition'':
$\partition_t \sim \text{LSP}(\bm{\mu}, \omega=3990.81)$ using
our permutation modification to make the analysis invariant to observation order,
ii. \mbox{``Centered Partition Process w/ VI Loss''}: $\partition_t \sim \text{CPP}(\bm{\mu}, \omega=30, \text{VI}, \text{CRP}(1))$,
iii. \mbox{``Centered Partition Process w/ Binder Loss''}: $\partition_t \sim \text{CPP}(\bm{\mu}, \omega=500, \text{Binder}, \text{CRP}(1))$,
iv. \mbox{``Shrinkage Partition -- Common''}: $\partition_t \sim \text{SP}(\bm{\mu}, \bm{\omega}, \psi=0.02, \text{CRP}(1))$
with $\bm{\omega} = (5, \ldots, 5)$, 
v.\ \mbox{``Shrinkage Partition -- Idiosyncratic''}: $\partition_t \sim \text{SP}(\bm{\mu}, \bm{\omega},\psi=0.02, \text{CRP}(1))$
with $\bm{\omega} = 5 \times (a_1, \ldots, a_n)$ in which $a_i = 1$ for all states except $a_i = 1/5$ for
Maryland, Delaware, District of Columbia, Montana, North Dakota, and South Dakota. These states
with $a_i = 1/5$ are on the borders of regions and it is reasonable to be less confident in the
allocation to their particular region in the anchor partition $\bm{\mu}$. 
Note that, for the sake of comparison, we fix the concentration 
parameter at $\alpha=1$ for the CRP baseline distribution.  In practice, one might treat $\alpha$ as random when using our SP 
distribution since it has a tractable normalizing constant.

To assess fit, we performed 10-fold cross validation as follows.  The dataset
was divided into 10 equally-sized, mutually-exclusive, and exhaustive shards (subsets).
For each fold, one of the 10 shards was held out as the test dataset for a model fit with
the other 9 shards.  For each fold, 55,000 MCMC iterations were run, the first 5,000 were
discarded as burn-in, and 1-in-10 thinning was applied for convenience.
For each MCMC iteration, the permutation $\perm_t$ received ten Metropolis update attempts based
on uniformly sampling ten items from $\perm_t$ and randomly shuffling them.
We computed the out-of-sample
log-likelihood, defined as the sum of the Monte Carlo estimate of the expectation of the log-likelihood contribution for each
observation when it was part of the held-out shard.  We then divide this sum by the sample size and multiply by 1,000,000 (for convenience) to get
a per-unit fit score, where larger values indicate better out-of-sample fit.
To aid in comparisons,
we show ``relative fit scores'' with reference to the $\text{CRP}(1)$, that is, the fit score for a particular model minus that of the $\text{CRP}(1)$.   Results
are summarized in Table \ref{tab:comparisons}, where the models are ordered with the 
best performing at the top.
Each of the SP, CPP, and LSP prior improve upon the one extreme of the CRP (which ignores the region information)
and the other extreme of fixing the partition at the anchor partition $\baselinepart$ defined by the regions.
In particular, the model with our SP prior performs 2,448 units better than the model using the CRP prior,
an improvement that we put in context in Section \ref{sec:temporal}.
Elapsed time for the full data analysis for each model using single-threaded code is included in the table.
All the timing metrics in this paper were computed on a server with 256 GB of RAM and with dual Intel Xeon Gold 6438Y+ CPUs.

\begin{table}[t]
\caption{Monte Carlo estimates of our relative fit score, defined as the mean out-of-sample log-likelihood (based on 10-fold cross validation) 
minus that of the $\text{CRP}(1)$, times 1,000,000.
Larger values indicate better fit.  The largest margin of error for 95\% Monte Carlo confidence intervals is 14 units.
The elapsed time for each model, in minutes, is also shown.}
\label{tab:comparisons}
\begingroup
\renewcommand{\arraystretch}{0.7} %
\begin{tabular}{lrS[table-format=3.1]}
\toprule
\textbf{Prior Partition Distribution} & \textbf{Relative Fit} & \textbf{Time} \\
\midrule
Shrinkage Partition Distribution -- Idiosyncratic    & 2,448 & 60.2 \\
Shrinkage Partition Distribution -- Common  & 1,854 & 60.3 \\
Centered Partition Process w/ VI Loss  & 1,842 & 32.6 \\
Location-Scale Partition & 1,272 & 97.1 \\
Centered Partition Process w/ Binder Loss  &  781 & 35.9 \\
Fixed Partition of $\baselinepart$  &  620 & 25.7 \\
Chinese Restaurant Process & 0 & 27.2 \\
Saturated Regression Model & -10,221 & \\
\bottomrule
\end{tabular}
\endgroup
\end{table}

\newcommand{\ols}{To consider a simple alternative model for the sake of benchmarking, Table
\ref{tab:comparisons} also contains the relative fit score of a saturated regression model having
covariates obtained by: i.\ interacting 27 year dummy variables with the nine
covariates in the $\bm{Z}$ matrix and ii.\ interacting the 27 year dummy
variables, 51 state dummy variables, and the four covariates in the $\bm{X}$
matrix.  (About a dozen three-way interactions could not be estimated due to limitations of the data.)
The saturated regression model does not permit the borrowing of strength (and assumes homoskedastic errors across years),
leading to  substantially worse out-of-sample performance.}\ols{}

\subsection{Application Using Hierarchically-Dependent Partitions}
\label{sec:hierarchical}

The model in the previous subsection assumed \textit{a priori} independence among the partitions 
$\bm{\pi}_1, \ldots, \bm{\pi}_T$.  Note that we are repeatedly clustering the same
items and one might expect an improvement by allowing the ``borrowing of strength'' in partition estimation.
In this subsection, we demonstrate the 
Bayesian hierarchical model presented in Section \ref{sec:hierarchical_model}.
This illustrates that, in a very intuitive way, dependent partition models can be formulated in the same way as 
other Bayesian hierarchical models.
We show that our model produces a large improvement in the fit score
when compared to the models in the previous subsection.

Our hierarchical model uses the sampling model as defined in (\ref{eq:like1}) and 
(\ref{eq:like2}) and a hierarchical prior for the partitions $\bm{\pi}_1, \ldots, \bm{\pi}_T$
given in (\ref{eq:hierarch}), with
$\bm{\mu} \sim \text{SP}(\bm{\rho}_{\text{r}}, \shrink_0, \perm_0, \psi_0, \text{CRP}(1))$,
$\shrink = \shrinkitem \times (1, \ldots, 1)$ and $\shrinkitem \sim \text{Gamma}(5,1)$,
$\perm_t$ having a discrete uniform distribution, $\psi$ having a uniform prior on the unit interval, and
baseline distribution $p_b$ being $\text{CRP}(1)$.
For the hyperparameters of the SP prior on $\bm{\mu}$, the anchor $\bm{\rho}_{\text{r}}$ is the partition defined by the US Census Bureau regions,
we set $\shrink_0 = \shrinkitem_0 \times (a_1, \ldots, a_n)$ using $a_i$ values from 
Section \ref{sec:single} and place a $\text{Gamma}(5,1)$ prior on $\shrinkitem_0$, we assume a uniform prior on $\perm_0$, and 
we place a uniform prior on the unit interval for the grit parameter $\psi_0$.
Of course, other distributional choices could be made for $p(\bm{\mu})$ and $\baselinedist$.
In any case, inference is straightforward 
since standard MCMC techniques are available when using the SP distribution due to its tractable 
normalizing constant.
These priors for the shrinkage and grit parameters were chosen as outlined
in Section \ref{sec:simulateprior} and demonstrated in Appendix \ref{sec:clusterestimation}.

The permutations were updated using a Metropolis step that was applied 10 times
per MCMC iteration for each permutation, with each randomly proposing to shuffle
5 items.  The acceptance rate for an individual update was 7\%.
The single-core runtime each Markov chain was approximately 103 minutes; about
52\% of that time was used to update the anchor partition, 3\% on the shrinkage
and grit parameters, 5\% on the permutation parameters, and the remaining 40\%
on the other year-specific parameters.

To compare models, we again use the same 10-fold out-of-sample fit measure with the same number of samples, burn-in, and thinning as in the previous subsection.
We repeated the whole exercise ten times with randomly selected starting values
and then combined the samples after burn-in.
The relative fit score was 3,539 (with margin of error 109 for a 95\% confidence interval), a substantial improvement of about 1,091 units from the best model fit
from the independence models in Section \ref{sec:single}.

\subsection{Application Using Temporally-Dependent Partitions}
\label{sec:temporal}

We demonstrated in Section \ref{sec:hierarchical} the hierarchically-dependent partition model defined in Section \ref{sec:hierarchical_model}.  
This model greatly improved results over the independent models in Section \ref{sec:single}, however, the hierarchical structure
did not capture the intuition that two adjacent years are likely more dependent than two years that are far apart. 
The data in this example are a time series for $t=1, \ldots, T$, and thus an autoregressive model in time 
for $\bm{\pi}_1, \ldots, \bm{\pi}_T$ is natural.
We now demonstrate that the temporal model in Section \ref{sec:temporal_model} improves upon the results 
of the hierarchical model.

We follow the outline of Section \ref{sec:temporal_model} and use
the same sampling model, priors on the shrinkage and grit parameters, 
number of posterior samples, burn-in, thinning, 10-fold cross validation, etc.\ as in Section \ref{sec:hierarchical}.
We assume
$\bm{\pi}_1 \sim \text{SP}(\bm{\rho}_{\text{r}}, \shrink_0, \perm_0, \psi_0, \text{CRP}(1))$,
with anchor $\bm{\rho}_{\text{r}}$ defined by the census regions,
$\shrink_0 = \shrinkitem_0 \times (a_1, \ldots, a_n)$ using $a_i$ values from 
Section \ref{sec:single} and placing a $\text{Gamma}(5,1)$ prior on $\shrinkitem_0$, a uniform prior on $\perm_0$, and 
a uniform prior on the unit interval for the grit parameter $\psi_0$.
The acceptance rate for a permutation update was 15\%.
The single-core runtime of each Markov chain was approximately 165 minutes; about
2\% of the that time was used to update the shrinkage
and grit parameters, 2\% on the permutation parameters, and the remaining 96\%
on the other year-specific parameters.

\begin{figure}[tb]
     \centering
     \begin{subfigure}[b]{0.47\textwidth}
         \centering
         \resizebox{0.95\linewidth}{!}{\includegraphics{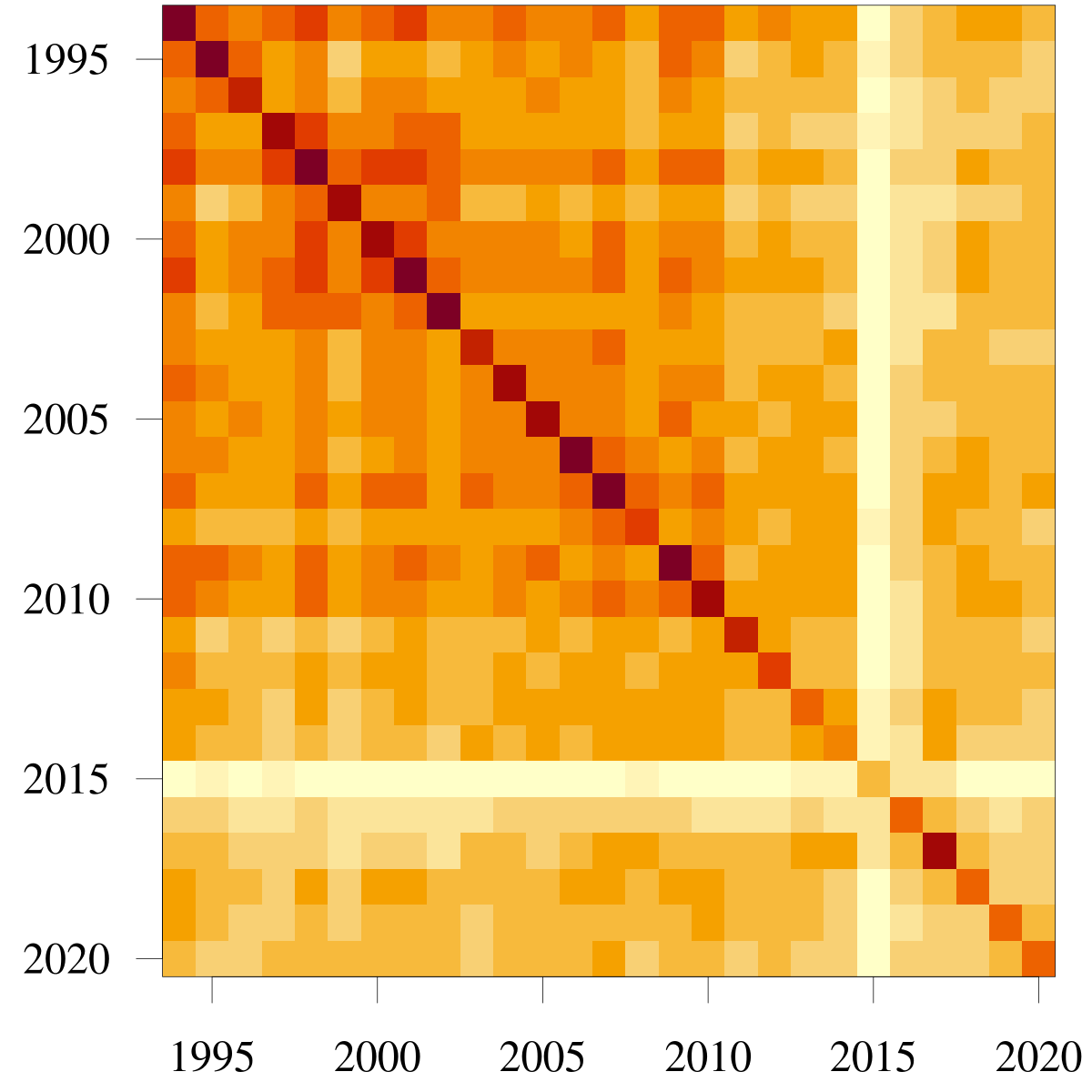}}
     \end{subfigure}
     \hfill
     \begin{subfigure}[b]{0.47\textwidth}
         \centering
         \resizebox{0.95\linewidth}{!}{\includegraphics{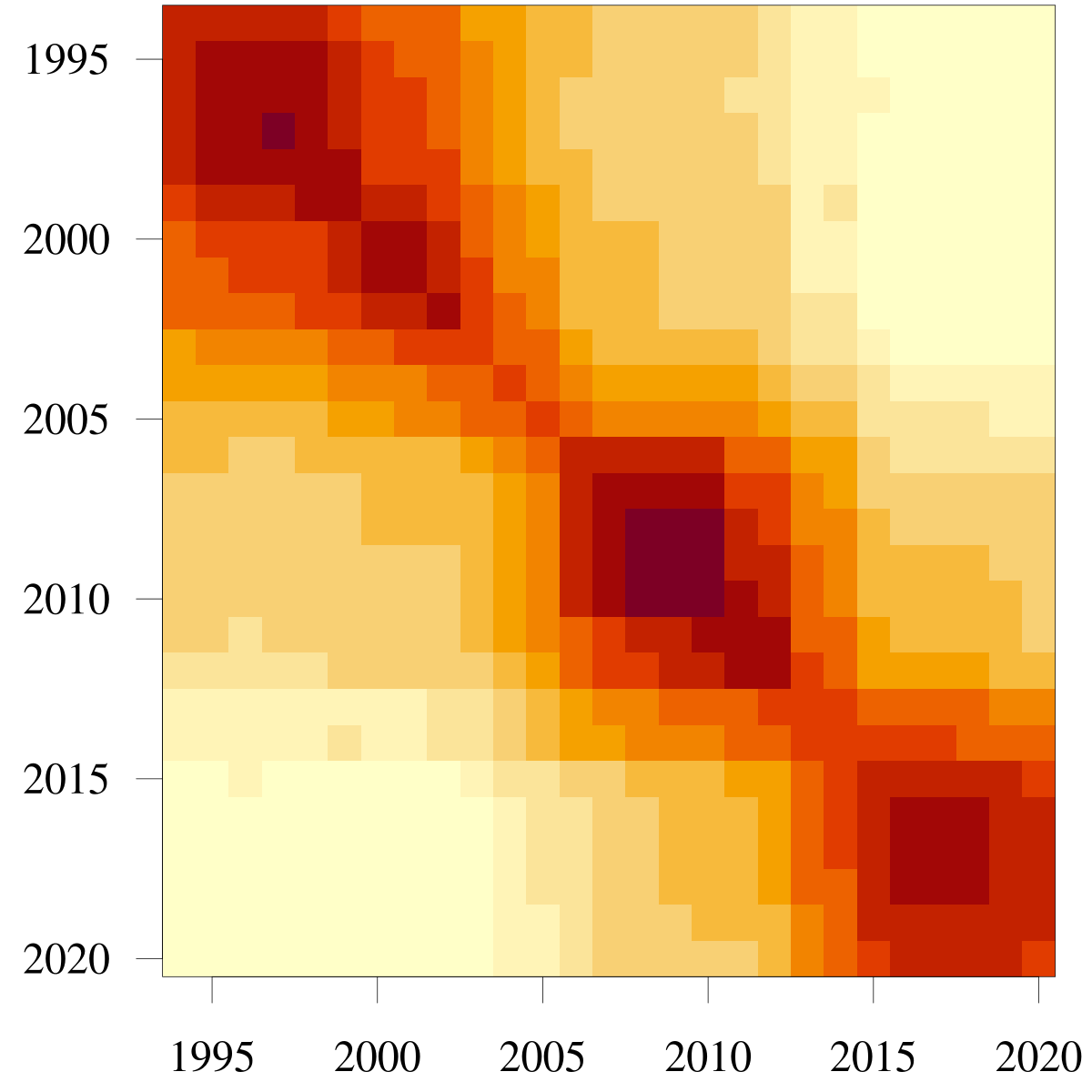}}
      \end{subfigure}
     \caption{The posterior expected Rand index for years 1994 to 2020.
     Darker colors indicate greater partition agreement. The left plot is for the independent model from Section
     \ref{sec:single} and the right plot is for our model for temporally-dependent random partitions.}
        \label{fig:MargLocScaleConst2}
\end{figure}

Figure \ref{fig:MargLocScaleConst2} helps visualize the dependence among the yearly partitions $\bm{\pi_1},\ldots,\bm{\pi_T}$.  
Each plot is a $27 \times 27$ grid with the $(i,j)$ cell showing the posterior expectation of
the Rand index (RI) between two randomly selected partitions from years $i$ and $j$.
More agreement in the partitions, i.e., larger RI value, are shown in darker colors in the plots.  
The left plot is for the independent CRP model in Section \ref{sec:single} and the right plot is for the time-dependent model.
There is a stark contrast between the plots of two models, with the temporally-dependent model demonstrating the anticipated temporal-dependence
in neighboring years which decays over time (i.e., moving away from the diagonal).

We again use the same 10-fold out-of-sample fit measure with the same number of samples, burn-in, and thinning
and repeated the whole exercise ten times.
The relative fit score was 5,000 $\pm$ 64, which is a substantial improvement of about 1,461 units from the hierarchical model in
Section \ref{sec:hierarchical}
and 2,552 units better than the best model fit
from the independence models in Section \ref{sec:single}.

\newcommand{\significance}{To explore the scientific significance of the
improvement in out-of-sample fit, we considered two approaches. First, we
randomly deleted 40\% and 50\% of the observations from each training dataset
in our 10-fold cross validation scheme, refit exactly the same model as
described in this section, and found that the relative fit score was 1,176
and -1,053, respectively.  We conclude that, for this study, one could reduce
the required sample size by about 45\% --- and therefore save significantly
in the cost of data acquisition --- while still achieving about the same
performance as the CRP(1) model with all the data. Second, we removed the
``married'' dummy variable, refit the model (with all of the observations),
and found the relative fit score was  1,026.  Likewise, we instead removed
both the ``white'' and ``Hispanic'' dummy variables, refit the model, and found
the relative fit was -1,780.  
We conclude that using our model with temporally-dependent partitions adds an approximate  
performance boost of an additional covariate 
to the CRP(1) model,  with the importance of that hypothetical covariate lying somewhere 
between ``married'' or the pair ``white'' and ``Hispanic''.
In practice, analysts would always
want to include these demographic dummy variables and would always want to
use all the available data.  This exercise merely demonstrates the scientific
significance of using our temporally-dependent partitions model over a simpler
model that does not capture partition dependence. Likewise, we conclude that
our hierarchical model and independent model using the SP distribution also both have
scientifically significant (albeit more modest) gains.}\significance{}

\section{Scalability}
\label{sec:scalability}

The added flexibility of the SP distribution in accommodating a ``best-guess''
partition or in modeling dependent partitions comes at a computational cost.
We first consider scalability in the number of time points $T$. We reran the
dependent models in Section \ref{sec:AppRTE} with data from 1994-2002 ($T =
9$), 1994-2011 ($T = 18$), 1994-2020 ($T=27$), 1994-2020 with 2020 down to 2011
($T=36$),  1994-2020 with 2020 down to 2002 ($T=45$), 1994-2020 with 2020 down
to 1994 ($T=54$). The elapsed time (using one CPU core) to fit the hierarchical
model ranged from 0.79 hours ($T=9$) to 2.89 hours ($T=54$), with a simple
linear regression (SLR) model in $T$ explaining almost $R^2 = 99\%$ of the
variation in the elapsed time. Likewise, the time to fit the temporal model
ranged from 1.14 hours ($T=9$) to 4.28 hours ($T=54$), with $R^2 = 98\%$.
Thus both the hierarchical and the temporal models appear to scale linearly with $T$.

We also investigated the scalability in the number of clusters.
Consider the data from 1994 and the model from Section \ref{sec:single}.  By
varying the concentration parameter $\alpha$ of the CRP($\alpha$) baseline
distribution in the SP distribution, we generated posterior samples for which
the average number of clusters ranged from 3.6 ($\alpha = 0.01$) to 14.4 ($\alpha
= 10$). The elapsed time was also recorded and a SLR of elapsed time on
the average number of clusters showed an adequate fit to the data ($R^2 = 95\%$),
suggesting that the elapsed time is approximately linear in the number of clusters.

While these models scale very well in the number of time points $T$ and number of
clusters, scaling in the number of items $n$ is more
challenging. Recall that all models to this point have clustered $n=51$
states.  To investigate scalability in $n$, we timed the SP model in Section
\ref{sec:single} using data from 1994 ($n=51$), data from 1994-1995 treating
states in different years as distinct items ($n=102$), data from 1994-1996
($n=153$), ..., data from 1994-2022 ($n=1,377$). The elapsed time (using one
CPU core) ranged from just over a minute ($n=51$) to just under 28 hours
($n=1,377$), and we found that the square root of the elapsed time
was linear in the number of items $n$, with a SLR having $R^2 = 99.9\%$.  Thus the SP model appears to scale quadratically in $n$.
In contrast, the elapsed time for the CRP model with $n=1,377$ was about
14 minutes.  Clearly the SP model is not practical for tens of thousands of
items, but we have demonstrated feasibility for at least $n = 1,377$ items.
\newcommand{\replusive}{We note, that one approach to reducing the computational cost may be to select a baseline distribution that favors a smaller number of clusters, for example, the CRP with a low mass parameter or a distribution with repulsive atoms similar to \cite{beraha2022mcmc}.}\replusive{}

\newcommand{\estimationupshot}{that the
SP distribution behaves as one would expected in a typical Bayesian model,
improving estimation when prior assumptions are met, worsening estimation
when prior assumptions are bad, and yielding to the data as the sample
size increases.}

\newcommand{\estimationcomparison}{conclude that the SP distribution
is competitive in cluster estimation while also being
amenable to posterior inference on
hyperparameters (due to its tractable normalizing constant) and ready-made for
dependent modeling (as demonstrated in Section \ref{sec:AppRTE}).}

\newcommand{\posteriorlearning}{show contours of the joint posterior density of the shrinkage $\shrinkitem$ and the grit $\psi$
for the first simulated dataset with $m=400$ --- the other 49 datasets look similar --- for the model
with $\bm{\mu} = \bm{\rho}_\text{r}$ (in solid curves) and $\bm{\mu} = \bm{\rho}
_\text{s}$ (in dashed curves). The movement from the prior mean
to the posterior means (indicated by the arrows) is also shown.  It is reassuring that the
posterior density moves toward higher shrinkage values when correctly assuming $\bm{\mu} = \bm{\rho}_\text{r}$
and toward smaller shrinkage values when mistakenly assuming $\bm{\mu} = \bm{\rho}_\text{s}$.
The separation of the prior and posterior densities indicate substantial posterior learning
on the shrinkage $\shrinkitem$ and grit $\psi$ parameters.}

\section{Discussion}
\label{sec:discussion}

We introduced a random partition distribution 
which shrinks a partition distribution $\baselinedist$ to an anchor 
partition $\baselinepart$ according to a shrinkage parameter $\shrink$.  
The primary motivation for this new partition distribution was to provide a 
straightforward mechanism to implement dependent partition models.  
We discussed and demonstrated the advantages of our SP distribution relative to the 
CPP and LSP.  We proved several intuitive properties of the SP distribution.
We then developed models for dependent partitions and 
demonstrated their performance in an empirical study.  We 
showed that adding dependency among related partitions, first using a hierarchical framework 
followed by time-series structure, improves the model performance quite dramatically. 
Appendix \ref{sec:clusterestimation} shows
\estimationupshot{}

\bibliographystyle{chicago}
\bibliography{main}

\newpage
\begin{appendices}
\setcounter{page}{1}

\section{An Extension to the Bell Number} \label{sec:eBells}

The $n^{\text{th}}$ Bell number provides the number of possible unique partitions of $n$ items.
Here we construct an extension of the Bell numbers enumerating how many unique
partitions are possible when allocating $a$ items to $b$ existing subsets.
We denote these numbers 
as $B(a,b)$ where $a$ and $b$ must be nonnegative integers.  We start with the trivial cases.  
First, by definition $B(0,0) \equiv 1$.  If there are $a$ items to allocate and $b=0$ (i.e., 
there are no existing subsets) then we have the standard Bell numbers $B(a,0)\equiv B(a)$.  
Another trivial case is when there are no items to allocate (i.e., $a=0$) with $b$ existing 
subsets then we have $B(0,b)\equiv 1$.  It is straightforward to see that $B(1,b) = b+1$, since 
there are $b$ current subsets and the one item to be allocated could go to the $b$ existing 
subsets or a new subset.

This extension to the Bell numbers also follow a similar recursive formula to that of the 
Bell numbers.  
We have:
\begin{equation} B(a,b+1) = \sum_{i=0}^{a} \genfrac(){0pt}{0}{a}{i} B(i,b). \end{equation}
To show this relationship is true we make the following argument.  First assume $B(a,b)$ is the 
number of possible unique partitions when allocating $a$ items and $b$ subsets already exist.  
We are now assuming we have $b+1$ existing subsets before allocating the $a$ items.  That extra subset must be contained in 
exactly one of the following cases:
\begin{itemize}
\itemsep0em
\item None of those $a$ items are added to that subset: there are $B(a,b)$ unique partitions in 
this case
\item One of those $a$ items are added to that subset: there are 
$\genfrac(){0pt}{1}{a}{1} B(a-1,b)$ unique partitions in this case 
\item Two of those $a$ items are added to that subset: there are 
$\genfrac(){0pt}{1}{a}{2} B(a-2,b)$ unique partitions in this case \\
\vdots
\item All $a$ of those items are added to that subset: there are 
$\genfrac(){0pt}{1}{a}{a} B(0,b)$ unique partitions in this case 
\end{itemize}
Recalling that $\genfrac(){0pt}{1}{a}{i} = \genfrac(){0pt}{1}{a}{a-i}$ and summing all the 
possible cases in reverse order, we have that 
$B(a,b+1) = \sum_{i=0}^{a} \genfrac(){0pt}{1}{a}{i} B(i,b)$.

A useful recurrence formula for these numbers are:
\begin{equation}\label{EQ:eBellRecur} B(a+1,b) = b \, B(a,b) + B(a,b+1). \end{equation}
We can show this formula is correct using the following logic.  Consider all possible ways to 
allocate the $(a+1)^{\text{th}}$ item (before the other $a$ items have been allocated to a 
subset).  There are two main cases:
\begin{itemize}
\itemsep0em
\item It goes to one of the existing $b$ subsets
\item It goes to a new singleton subset
\end{itemize}
If the $(a+1)^{\text{th}}$ item goes to one of the existing subsets then there are $B(a,b)$ ways 
to allocate the other items.  However, we need to multiply that by the number of ways the 
$(a+1)^{\text{th}}$ item can be allocated to the preexisting subsets.  Thus for the first case we 
have $b \, B(a,b)$ possible partitions.

If the $(a+1)^{\text{th}}$ item goes to a new singleton subset, then there are $b+1$ existing 
subsets in which to allocate $a$ items, or $B(a,b+1)$.  These two cases comprise all 
possibilities and give the number for adding a new item to allocate.

Note that (\ref{EQ:eBellRecur}) is especially useful in generating these numbers.

\newpage{}
\section{Proofs and Properties}
\label{apx:blc}

\subsection{Proof of Theorem \ref{thm:BLC}}
{\bf Theorem} \ref{thm:BLC}:
\textit{
\PropOne
}

\noindent {\bf Proof:}

\noindent For {\bf part a)} this property is easily demonstrated by looking at the CAPF of the SP distribution and setting $\shrinkitem=0$.  Now the exponential term is equal to 1 and we are left with the CAPF of $\baselinedist$.

\noindent For {\bf part b)} under the conditions that $\partition=\baselinepart$, 
$\shrink = \shrinkitem \times (1, \dots, 1)$, and $0<\shrinkitem<\infty$, the CAPF of the SP distribution reduces to:
\begin{align} \label{eq:pmfreduce}
\text{Pr}&_\text{sp}(\partitionitem_{\permitem_k} = \baselineitem_{\permitem_k} \mid \partitionitem_{\permitem_1}, \ldots, 
\partitionitem_{\permitem_{k-1}}, \baselinepart, \shrinkitem, \perm, \psi, \baselinedist ) 
= \\
&\frac{ \text{Pr}_\text{b}(\partitionitem_{\permitem_k} = \baselineitem_{\permitem_k} \mid \partitionitem_{\permitem_1}, \ldots, 
\partitionitem_{\permitem_{k-1}}) \, \times \exp \left(\frac{(1-\psi)\,\shrinkitem^2}{k-1} \sum_{j=1}^{k-1} \text{I} \{\baselineitem_{\permitem_j} = \baselineitem_{\permitem_k}\}  \right) }{
\displaystyle \sum_{c \, \in \, \mathcal{S} }\text{Pr}_\text{b}(\partitionitem_{\permitem_k} = c \mid \partitionitem_{\permitem_1}, \ldots, 
\partitionitem_{\permitem_{k-1}}) \, \times \exp \left(\frac{( \text{I}\{\baselineitem_{\permitem_k} = c \}-\psi)\,\shrinkitem^2}{k-1} \sum_{j=1}^{k-1} \text{I} \{\baselineitem_{\permitem_j} = c\} \right)
} \notag
\end{align}
for $k = 2,\ldots,n$ and where $\mathcal{S} = \{\baselineitem_{\permitem_1},\ldots,\baselineitem_{\permitem_{k-1}}, | \{\baselineitem_{\permitem_1},\ldots,\baselineitem_{\permitem_{k-1}} \}| +1 \}$.
This can be justified since if $\partition=\baselinepart$ 
then $\text{I} \{\partitionitem_{\permitem_j} = c\} = \text{I} \{\baselineitem_{\permitem_j} = c\}$. Further, 
in the numerator, since
$c=\baselineitem_{\permitem_k}$ we have $\text{I} \{\baselineitem_{\permitem_j} = c\}=\text{I} \{\baselineitem_{\permitem_j} = \baselineitem_{\permitem_k}\}$.

We are given that $\psi$ is confined to the interval $(0,1)$.  
For a fixed value of $k$, if $c$ agrees with $\baselinepart$ (i.e., $\text{I} \{\baselineitem_{\permitem_j} = c\} = 1$) 
then the exponential term is positive and goes to infinity with $\shrinkitem$.  
If $c$ does not agree with $\baselinepart$ (i.e., $\text{I} \{\baselineitem_{\permitem_j} = c\} = 0$) 
then the exponential term is negative and goes to 0 as $\shrinkitem \to \infty$.  
Thus for each $k$ the CAPF forces the allocation to 
agree with $\baselinepart$ with probability 1 and, using the algebraic limit theorem, we 
have $\lim_{\shrinkitem \to \infty} \text{Pr} \,(\partition=\baselinepart) = 1$.

\noindent For {\bf part c)}
we know $\psi \in (-\infty,0)$.  The expression in the exponential part of the SP's CAPF of (\ref{Eq:SPAnchorParts}) can be expressed as:
\begin{equation} \label{eq:CAPFexpPart}
 \frac{\shrinkitem^2}{k-1} \sum_{j=1}^{k-1}  \text{I}\{ 
\partitionitem_{\permitem_j} = c \} \left(  \text{I}\{ \baselineitem_{\permitem_j} = \baselineitem_{\permitem_k} \} - \psi \right).  
\end{equation}
The value of $(\text{I}\{ \baselineitem_{\permitem_j} = \baselineitem_{\permitem_k} \} - \psi)$ will always be positive in this case.  
Since by definition $\partitionitem_{\permitem_1} = 1$, the second item will be forced to be allocated with the first, i.e., Equation (\ref{eq:CAPFexpPart}) will be infinite if $\partitionitem_{\permitem_2}=1$ and zero otherwise.  The same will hold for each subsequent item allocation.
Thus if $\psi \in (-\infty, 0)$ then 
$\lim_{\shrinkitem \to \infty} Pr(\partition=\bm{\rho}_1) = 1$.

\noindent For {\bf part d)}
we make a similar argument as in part c).  If $\psi \in (1,\infty)$ then (\ref{eq:CAPFexpPart}) will always be zero or negative.  A partition that forces this part of the CAPF to always be zero will have positive probability mass as $\shrinkitem$ gets large.  The partition $\bm{\rho}_n$ (each item is allocated to a unique cluster) is the only partition that matches that criterion since $\text{I}\{ 
\partitionitem_{\permitem_j} = c \}$ will always be equal to 0.  As items are sequentially allocated each item must go to its own cluster with probability 1, and no other partition has that property.   Thus if $\psi \in (1,\infty)$ then $\lim_{\shrinkitem \to \infty} Pr(\partition=\bm{\rho}_n) = 1$. \qed

In general, we recommend specifications of the SP distribution where $\psi \in (0,1)$ because the distribution is consistent with the anchor partition as demonstrated in {\bf part b}. This is an ideal property for applications where the anchor partition is elicited as the prior clustering configuration and the intended interpretation of the shrinkage $\omega$ is a degree of compromise between the baseline $p_b$ and the anchor partition $\bm\rho$. While $\psi \notin (0,1)$ is a valid specification as implied in {\bf part c} and {\bf part d}, we urge caution interpreting the SP distribution parameters when consistency does not hold.

\subsection{Proof of Theorem \ref{thm:monotonic_pm}}
{\bf Theorem} \ref{thm:monotonic_pm}:
\textit{
\theoremOne
}

\noindent {\bf Proof:}

\noindent Under the conditions that $\partition=\baselinepart$, 
$\shrink = \shrinkitem \times (1, \dots, 1)$ and $0<\shrinkitem<\infty$, the CAPF of the SP distribution reduces to (\ref{eq:pmfreduce}).
To streamline notation we let $d_k(c) = \sum_{j=1}^{k-1} \text{I} (\baselineitem_{\permitem_j}=c)/(k-1)$ and write $\text{Pr}_\text{b}(\partitionitem_{\permitem_k} = c \mid \partitionitem_{\permitem_1}, \ldots, 
\partitionitem_{\permitem_{k-1}})$ as $\text{Pr}_\text{b}(\partitionitem_{\permitem_k} = c \mid \cdot \, )$.  Thus the CAPF can be more compactly written as:
\begin{align*} 
\begin{split}
&\frac{ \text{Pr}_\text{b}(\partitionitem_{\permitem_k} = \baselineitem_{\permitem_k} \mid \cdot \, ) \, \exp \left(d_k(\baselineitem_{\permitem_k}) \, (1-\psi)\,\shrinkitem^2 \right) }{
\displaystyle \sum_{c \, \in \, \mathcal{S} }\text{Pr}_\text{b}(\partitionitem_{\permitem_k} = c \mid \cdot \, ) \,  \exp \left(d_k(c) \, ( \text{I}\{\baselineitem_{\permitem_k} = c \}-\psi)\,\shrinkitem^2 \right)
}.
\end{split}
\end{align*}

We want to show:
\begin{align*}
&\text{Pr}_\text{sp}(\partitionitem_{\permitem_k} = \baselineitem_{\permitem_k} \mid \partitionitem_{\permitem_1}, \ldots,\partitionitem_{\permitem_{k-1}}, \baselinepart, \shrinkitem + \delta, \perm, \psi, \baselinedist ) \\ & \geq  \ \text{Pr}_\text{sp}(\partitionitem_{\permitem_k} = \baselineitem_{\permitem_k} \mid \partitionitem_{\permitem_1}, \ldots, \partitionitem_{\permitem_{k-1}}, \baselinepart, \shrinkitem, \perm, \psi, \baselinedist )
\end{align*}
for $k=2,\ldots,n$ and that the inequality is strict for at least one $k$ and $\perm$.  This is equivalent to showing:
\begin{align*} 
\begin{split}
&\displaystyle \sum_{c \, \in \, \mathcal{S} }\text{Pr}_\text{b}(\partitionitem_{\permitem_k} = c \mid \cdot \, ) \, \exp \left(d_k(c) \, ( \text{I}\{\baselineitem_{\permitem_k} = c \}-\psi)\,\shrinkitem^2 \right) \, \exp \left(d_k(\baselineitem_{\permitem_k}) \, (1-\psi)\, (\shrinkitem+\delta)^2 \right)\\
 & \geq  \
\displaystyle \sum_{c \, \in \, \mathcal{S} }\text{Pr}_\text{b}(\partitionitem_{\permitem_k} = c \mid \cdot \, ) \, \exp \left(d_k(c) \, ( \text{I}\{\baselineitem_{\permitem_k} = c \}-\psi)\,(\shrinkitem+\delta)^2 \right) \, \exp \left(d_k(\baselineitem_{\permitem_k}) \, (1-\psi)\, \shrinkitem^2 \right)
\end{split}
\end{align*}
Before we drop the $\text{Pr}_\text{b}(\partitionitem_{\permitem_k} = c \mid \cdot \, )$ from consideration, recall that $0<\baselinedist(\baselinepart)<1$, thus there must exist at least one $\perm$, $k$, and $c \neq \partitionitem_{\permitem_k}$ such that $\text{Pr}_\text{b}(\partitionitem_{\permitem_k} = c \mid \cdot \, ) > 0$.
Now comparing the summands for each possible $c$, we see if $c=\baselineitem_{\permitem_k}$ then the terms on each side of the inequality are equal.  If $c \neq \baselineitem_{\permitem_k}$ then we can reduce the problem to:
\begin{align*} 
\begin{split}
 - d_k(c) \, \psi \,\shrinkitem^2 + d_k(\baselineitem_{\permitem_k}) \, (1-\psi)\, (\shrinkitem+\delta)^2
  \geq  \
- d_k(c) \, \psi \,(\shrinkitem+\delta)^2 + d_k(\baselineitem_{\permitem_k}) \, (1-\psi)\, \shrinkitem^2.
\end{split}
\end{align*}
This is clearly true if $\psi \in (0,1)$, however, for at least one $c\in \mathcal{S}$ we can impose a strict inequality since both $d_k(c) = 0$ and $d_k(\baselineitem_{\permitem_k}) = 0$ cannot occur simultaneously for all $c$ not equal to $\baselineitem_{\permitem_k}$.  Thus: 
\begin{align*}
&\text{Pr}_\text{sp}(\partitionitem_{\permitem_k} = \baselineitem_{\permitem_k} \mid \partitionitem_{\permitem_1}, \ldots,\partitionitem_{\permitem_{k-1}}, \baselinepart, \shrinkitem + \delta, \perm, \psi, \baselinedist ) \\ & >  \ \text{Pr}_\text{sp}(\partitionitem_{\permitem_k} = \baselineitem_{\permitem_k} \mid \partitionitem_{\permitem_1}, \ldots, \partitionitem_{\permitem_{k-1}}, \baselinepart, \shrinkitem, \perm, \psi, \baselinedist )
\end{align*}
for all $k = 2, \ldots, n$.
Since the inequality is true for an arbitrary $\perm$ and strict for at least one, it also strict when $\perm$ is integrated out of the model, therefore 
$\text{Pr}\,(\partition_1 = \baselinepart) < \text{Pr}\,(\partition_2 = \baselinepart)$.\qed

\subsection{Proof of Theorem \ref{thm:kullback}}
{\bf Theorem} \ref{thm:kullback}:
\textit{
\theoremTwo
}

\noindent {\bf Proof:}

\noindent For the Kullback-Leibler divergence:

Let $\Pi$ be the space of all partitions (of $n$ items) and $0 \leq \shrinkitem < \shrinkitem +\delta \leq \infty$.  
Then the Kullback-Leibler divergence between $\baselinepart \sim p_{\baselinepart}$ and 
$\partition_1 \sim \text{SP}(\baselinepart, \shrink, \psi, \baselinedist)$ is:
\begin{align*}
D_{KL}\big(\baselinepart, \partition_1 
\big) &= \sum_{\partition \in \bm{\Pi}} p_{\baselinepart}
(\partition)\log\left(\frac{p_{\baselinepart}(\partition)}{\shrinkdist(\partition \mid 
\baselinepart, \shrink, \psi, \baselinedist)}\right)\\
&= 1 \cdot \log
\left(\frac{1}{\shrinkdist(\baselinepart \mid \baselinepart, \shrink, \psi,  
\baselinedist)}\right)\\
&= - \log\big(\shrinkdist(\baselinepart \mid \baselinepart, \shrink, \psi, 
\baselinedist)\big)
\end{align*}
By Theorem \ref{thm:monotonic_pm} we know $
    \shrinkdist(\baselinepart \mid \baselinepart, \shrink, \psi, \baselinedist) < 
    \shrinkdist(\baselinepart \mid \baselinepart, \shrink + \delta, \psi, 
    \baselinedist)$.
Transforming this equation we have 
$-\log(\shrinkdist(\baselinepart \mid \baselinepart, \shrink, \psi,  \baselinedist)) > 
-\log(\shrinkdist(\baselinepart \mid \baselinepart, \shrink + \delta, \psi, 
\baselinedist))$.  Thus if $\partition_2 \sim {SP}(\baselinepart, \shrink+\delta, \psi, \baselinedist)$, then $D_{KL}\big(\baselinepart, \partition_2 \big) < D_{KL}\big(\baselinepart, \partition_1 \big)$. \qed \\

\noindent For the total variance distance:

We define the total variation distance between $\baselinepart \sim p_{\baselinepart}$ and 
$\partition_1 \sim \text{SP}(\baselinepart, \shrink, \psi, \baselinedist)$ to be:
\[ D_{TV}\big( \baselinepart, \partition_1 \big) =  \frac{1}{2}\sum_{\partition \in \Pi} \, \Big|  \,
p_{\baselinepart}(\partition) -  p_{\text{sp}}(\partition \mid \baselinepart, 
\shrink, \psi, \baselinedist) \,\Big| \,. \]
This reduces to $1-p_{\text{sp}}(\baselinepart \mid \baselinepart, \shrink, \psi, \baselinedist)$, since $p_{\baselinepart}(\partition)$ is always zero except when $\partition=\baselinepart$.  
By Theorem \ref{thm:monotonic_pm},
$p_{\text{sp}}(\baselinepart \mid \baselinepart, \shrink, \psi, \baselinedist) < p_{\text{sp}}
( \baselinepart \mid \baselinepart, \shrink + \delta, \psi, \baselinedist)$.
Thus if $\partition_2 \sim \text{SP}(\baselinepart, \shrink+\delta, \psi, \baselinedist)$, then $D_{TV} \big( \baselinepart, \partition_2 \big) < D_{TV} \big( \baselinepart, \partition_1 \big)$. \qed \\

\noindent For the expected Rand index converging to 1:

From Theorem \ref{thm:monotonic_pm} we know $P(\partition=\baselinepart)=1$ as $\shrinkitem \to \infty$. If $\partition=\baselinepart$ with probability 1 then it is equivalent to show that the Rand index is equal to 1 with probability 1 or $P(RI(\baselinepart,\partition)=1)=1$.  Clearly $P(RI(\baselinepart,\partition)=1) = P(RI(\baselinepart,\partition) \geq 1)$, then by the Markov inequality $P(RI(\baselinepart,\partition) \geq 1) \leq E_{\baselinepart,\partition}[RI(\baselinepart,\partition)]$.  Thus in the limit as $\shrinkitem \to \infty$, $E_{\baselinepart,\partition}[RI(\baselinepart,\partition)]=1.$
\qed

\subsection{Proof of Theorem \ref{thm:zeroweights}}
{\bf Theorem} \ref{thm:zeroweights}:
\textit{
\theoremThree
}

\noindent {\bf Proof:}

\noindent To prove this theorem consider an arbitrary fixed partition $\bm{\rho}$. We show that  under 
the conditions of the theorem $\text{Pr}(\partition_1=\bm{\rho}) = \text{Pr}(\partition_2=\bm{\rho})$, 
thus implying $\partition_1$ is equal in distribution to $\partition_2$.

Without loss of generality assume the items will be allocated using the natural permutation. From the 
CAPF in (\ref{Eq:SPAnchorParts}), with the permutation parameter removed,
it is evident that both 
$\baselinepart$ and $\shrink$ only influence the exponential term.  That exponential term is:
\[
\exp \left( \frac{\shrinkitem_{i}}{i-1}   \sum_{j=1}^{i-1}  \shrinkitem_{j} \, \text{I}\{ 
\partitionitem_{j} = c \} \left( \text{I}\{ \baselineitem_{j} = \baselineitem_{i} \}  - \psi \right)
\right).
\]
For each $i \in \{2,\ldots,n\}$ it is clear that:
\begin{align*}
    &\exp \left( \frac{\shrinkitem_{i}}{i-1}   \sum_{j=1}^{i-1}  \shrinkitem_{j} \, \text{I}\{ \partitionitem_{j} = c \} \left( \text{I}\{ \baselineitem_{j} = \baselineitem_{i} \}  - \psi \right)
\right) \\
= \ & 
\exp \left( \frac{\shrinkitem_{i}}{i-1}   \sum_{j=1}^{i-1}  \shrinkitem_{j} \, \text{I}\{ 
\partitionitem_{j} = c \} \left( \text{I}\{ \baselineitem_{j}^* = \baselineitem_{i}^* \}  - \psi \right)
\right)
\end{align*}
since any time $\baselineitem_{j} = \baselineitem_i$ 
does not imply that $\baselineitem^{*}_j = \baselineitem^{*}_j$ (or vice versa)
we have that either $\shrinkitem_i=0$ or $\shrinkitem_j=0$.
Thus the CAPFs of $\partition_1$ and $\partition_2$ 
produce the same probabilities for an arbitrary 
partition.  
Additionally, the above equalities are not dependent on the allocation order of the items, and they 
also hold for any $\perm$.  Therefore the pmfs for $\partition_1$ and $\partition_2$ produce the same 
probabilities for an arbitrary permutation and are equal in distribution.  \qed

\subsection{Proof of Theorem \ref{prop:limiting}}
{\bf Theorem} \ref{prop:limiting}
\textit{
\propTwo
}

\noindent {\bf Proof:}
The crux of the proof is to show under the theorem's conditions and
with $\partition \sim \text{SP}(\baselinepart,\shrink,\perm,\psi,\baselinedist)$ then:
\[\lim_{\shrinkitem \to \infty} \text{Pr}(\partition = \bm{\rho}) > 0.\]
First consider a restricted permutation $\perm^*$, such that all items with $s_i=1$ are allocated first, and the items with $s_i=0$ are allocated afterwards. 
For the items with $s_i=1$, any time 
$\baselineitem_{\permitem_i}=\baselineitem_{\permitem_j}$ we have that 
$\rho_{\permitem_i}=\rho_{\permitem_j}$.  Therefore $\bm{\rho}$ has a probability of 1 up to that point in the allocation (see Theorem \ref{thm:BLC}).  Now as the final 
items with $s_i=0$ are allocated each of their CAPFs must also be nonzero since 
$\baselinedist(\bm{\rho})>0$.  Thus for any permutation $\perm^*$ in that restricted class
\[\lim_{\shrinkitem \to \infty} \text{Pr}(\partition = \bm{\rho}) > 0.\]
Now consider an arbitrary permutation $\perm$.  If the allocation order of an item with $s_i=0$ is moved before an item with $s_i=1$, we have the following:
\begin{enumerate}
\item The exponential term in the SP’s CAPF
is simply 1 and the item is allocated using the probabilities of $\baselinedist$, which still must be positive.
\item Additionally, for the item with $s_i=1$, the sums in its exponential term of the SP's CAPF, (\ref{Eq:SPAnchorParts}), are unchanged whether or not an item with $s_i=0$ is allocated before it or not.  These sums produced a positive exponential term in CAPF for the case with the restricted $\perm^*$, thus they must also produce a positive CAPF value when considering an arbitrary $\perm$ (with the condition that $\baselinedist(\bm{\rho})>0$). 
\end{enumerate}
These two arguments can be iterated any number of times to produce any $\perm$ by permuting the order of $\perm^*$ one item at a time.  Therefore we have that $\bm{\rho}$ is a limiting partition of $\text{SP}(\baselinepart,\shrink,\psi,\baselinedist)$.  \qed

\subsection{Proof of Theorem \ref{prop:NumLimParts}}
{\bf Theorem} \ref{prop:NumLimParts}
\textit{
\propThree.
}

\noindent {\bf Proof:}
There are $n$ items being partitioned, with $a$ items that have an idiosyncratic 
shrinkage parameter of zero.  There are also $b$ clusters in $\baselinepart$ that have least 1 
item such that $\shrinkitem_{i} > 0$.  

Without loss of generality, let the $n-a$ items with $\shrinkitem_{i} > 0$ be 
allocated first.  By Theorem \ref{thm:BLC}, with probability 1, they must be allocated to agree 
with $\baselinepart$ (since $\shrinkitem \to \infty$).  Thus there is only one possible 
partition for those $n-a$ items.  Now if the other $a$ items are allocated, there are exactly 
$b$ existing clusters and $a$ items to be allocated, thus there are exactly $B(a,b)$ possible 
partitions (see Appendix \ref{sec:eBells}).  The item allocation order can change the 
probability of a possible partition, but in this case it cannot change the number of 
possibilities that have probability greater than zero.  \qed

\newpage{}
\section{Details of the MCMC Algorithms} \label{sec:MCMC_Algo}

This appendix provides details of the MCMC algorithms used in Section
\ref{sec:AppRTE}.
Software implementing our SP distribution is available as an
R package (\url{https://github.com/dbdahl/gourd-package}).
Table \ref{tab:notation} reiterates notation.   Section \ref{sec:modAlgo8} introduces the
most intricate aspect of the MCMC involving our SP distribution, which
is updating the  random partition based on a modification of Neal's
Algorithm 8 \citep{neal2000markov}. Then, Sections \ref{sec:indMCMC} -
\ref{sec:temporalMCMC} explain the sampling details use for the independent
model in Section \ref{sec:single}, the hierarchical model in Section
\ref{sec:hierarchical}, and the temporal model in  Section \ref{sec:temporal},
respectively.

\begin{table}[H]
\caption{Notational summary.  Lower case bolded items  
are vectors, and capitalized bolded items are matrices.}
\label{tab:notation}
\begingroup
\fontsize{10pt}{10pt}\selectfont
\renewcommand{\arraystretch}{0.7} %
\begin{tabularx}{450pt}{cX}
\toprule
    \textbf{Symbol} & \textbf{Description} \\
\midrule
    $n$ & Number of states in the US (including the  District of Columbia), i.e., $n = 51$. \\
    $T$ & Total number of years under consideration, i.e., $T = 27$.\\
    $\bm{I}$ & An identity matrix.\\
    $n_{it}$ & Number of observations in the $i^{\text{th}}$ state and $t^{\text{th}}$ year. \\ 
    $n_{ct}$ & Sum  of the $n_{it}$ of each state for the $t^{\text{th}}$ year and in cluster $c$.\\
    $\bm{y}_{t}$ & Log hourly earnings for individuals in the $t^{\text{th}}$ year, length $\sum_{i=1}^{n}n_{it}$.\\
    $\bm{y}_{ct}$ & Log hourly earnings for individuals in the $c^{\text{th}}$ cluster and $t^{\text{th}}$ year, length $n_{ct}$.\\
    $\bm{y}_{it}$ & Log hourly earnings for individuals in the $i^{\text{th}}$ state and $t^{\text{th}}$ year, length $n_{it}$.\\
    $\bm{Z}_{t}$ & Demographic covariates for individuals in the $t^{\text{th}}$ year, dimension $\sum_{i=1}^{n}n_{it} \times 9$.\\
    $\bm{Z}_{ct}$ & Demographic covariates for the $c^{\text{th}}$ cluster and $t^{\text{th}}$ year, dimension $n_{ct} \times 9$.\\
    $\bm{Z}_{it}$ & Demographic covariates for the $i^{\text{th}}$ state and $t^{\text{th}}$ year, dimension $n_{it} \times 9$.\\
    $\bm{\gamma}_t$ & Common regression coefficients for the $t^{\text{th}}$ year associated with $\bm{Z}_{t}$, length 9.\\
    $\bm{X}_{ct}$ & Intercept and education dummy variables for the $c^{\text{th}}$ cluster and $t^{\text{th}}$ year, dimension $n_{ct} \times 4$.\\ 
    $\bm{X}_{it}$ & Intercept and education dummy variables for the $i^{\text{th}}$ state and $t^{\text{th}}$ year, dimension $n_{it} \times 4$.\\ 
    $\bm{\beta}^*_{ct}$ & Coefficients for intercept and education dummy variables associated with $\bm{X}_{ct}$, length 4.\\
    $\bm{\beta}_{it}$ & Coefficients for intercept and education dummy variables associated with  $\bm{X}_{it}$, length 4.\\
    $\tau_t$ & Model precision for the $t^{\text{th}}$ year.\\
    $p_{\text{sp}}$ & Probability mass function (pmf) of the SP distribution, given in (\ref{Eq:SPDpmfSecond-old}).\\
    $\partition_t$ & Partition cluster labels in canonical form for the $t^{\text{th}}$ year, length $n$.\\
    $q_t$ & Number of clusters in a given $\partition_t$.\\
    $\baselinepart$ & Cluster labels in canonical form, defining the anchor for all years, length $n$.\\
    $\baselinepart_t$ & Cluster labels in canonical form, defining anchor for the $t^{\text{th}}$ year, length $n$.\\
    $\shrink$ & The shrinkage parameter in the SP distribution, length $n$.\\
    $\perm_t$ & Permutation of $1,2,\ldots,n$ used to define the order of item allocation for $\partition_t$, length $n$.\\
    $\bm{\rho}_{\text{r}}$ & A fixed partition defined by the US Census Regions as shown in Figure \ref{fig:regions}, length $n$.\\
    $\bm{\rho}_{\text{s}}$ & A fixed partition defined by a random shuffle of $\bm{\rho}_{\text{r}}$, length $n$.\\
    $\psi$ & Grit parameter in the SP distribution, typically in $(0,1)$.\\
    $\baselinedist$ & Baseline partition distribution of the SP distribution.\\
\bottomrule
\end{tabularx}
\endgroup
\end{table}

\subsection{Updating the Partition} \label{sec:modAlgo8}
Our MCMC update of a partition with an SP prior follows a modification of
Neal's Algorithm 8, where each step updates a cluster label of a given item conditional on the allocation of
all other items.
At each step, we assume for notational convenience that the partition $\partition_t$ is represented in 
cluster label notation using a canonical form, i.e., each 
element of the vector has a value $c=1,\ldots,q_t$ and, 
according to the permutation, 1 must appear before 2, 2 
before 3, etc.  For a fixed year $t$ (where $t=1,\ldots,27$) and $i$ (where 
$i=1,\ldots,n$), update the cluster label of the 
${\permitem_{ti}}^{\text{th}}$ item using the following 
steps:
\begin{enumerate}
\item Define $q_{t}$ to be the current number of  clusters in $\partition_t$,
let $\partition_t^*$ to be identical to $\partition_t$ with the exception that $\partitionitem_{t\permitem_{t i}}^* = c$, and 
let $\bm{y}_{it}\mid \bm{\beta}^*_{ct},\bm{\gamma}_t,\tau_t \sim \text{N}(\bm{X}_{it}\bm{\beta}^*_{ct}+
\bm{Z}_{it}\bm{\gamma}_t, \tau_t \bm{I})$, i.e., $p(\bm{y}_{it}\mid \bm{\beta}^*_{ct},\bm{\gamma}_t,\tau_t)$ is a 
multivariate normal density with precision $\tau_t \bm{I}$.
Then for $c=1,\ldots,q_{t}$, calculate
\begin{equation} \label{eq:algo8probs1}
    p(\bm{y}_{it} \mid \bm{\beta}^*_{ct},\bm{\gamma}_t,\tau_t) \times p_\text{sp}(\partition_t^* \mid \baselinepart, \shrink, \perm_t, \psi, \baselinedist). 
\end{equation}

\item If the ${\permitem_{ti}}^{\text{th}}$ item is not currently in a singleton cluster, draw a value of 
$\bm{\beta}^*_{ct}$ from a $\text{N}\left(\bm{\mu}_{\beta}, 
\tau_{\beta} \bm{I} \right)$ and once again calculate 
(\ref{eq:algo8probs1}), where $c = q_t+1$.  That is, we only propose one possible new 
cluster (i.e., $m=1$ in the notation of Neal's Algorithm 8).  It appears to
be straightforward to allow for the $m>1$ case, if desired. 

\item With probabilities proportional to the quantities 
calculated in the previous two steps, randomly select which
cluster the ${\permitem_{ti}}^{\text{th}}$ item 
is reassigned. We note that there are
computational efficiencies in (\ref{eq:algo8probs1}) and 
subsequent equations since the CAPF terms for items 
$1,\ldots,i-1$ are the same for all possible clusters $c$ 
and would therefore cancel in the quantities calculated in this 
step.   
\end{enumerate}
One iteration of the algorithm involves repeating the previous steps for each of the $n$ times.

\subsection{Independent Models} \label{sec:indMCMC}
We now describe the details of the MCMC scheme 
for the SP models of Section \ref{sec:single}.
Although there are actually two SP
models, they only differ in their fixed value of $\shrink$.
Each step of the MCMC is the same, thus we will refer to 
it as a single model.

The details of how the parameters are updated, and 
initialized, are given below.  The numbered ordering below
is arbitrary, these could be rearranged without any 
impact to the scheme's validity.  

\begin{enumerate}

\item $\partition_t$: The prior for each 
$\partition_t \mid 
\baselinepart_t,\shrink,\perm_t,\psi,\baselinedist$ 
is an independent $\text{SP}
(\baselinepart_t,\shrink,\perm_t,\psi,\baselinedist)$. 
This is a one-item-at-a-time Gibbs update. For each year 
$t$, update the partition according to the steps in Section
\ref{sec:modAlgo8}.    The value of $\partition_t$ is 
initialized at the fixed value of $\bm{\rho}_{\text{r}}$ 
and $\baselinedist$ is the CRP(1); the other parameters are
described below.

\item $\baselinepart_t$: In this model, the anchor for each
year is fixed at $\bm{\rho}_{\text{r}}$, so no MCMC updates
are needed.   

\item $\perm_t$: The prior for each $\perm_t$ is a discrete
uniform over the space of permutations (\textit{iid} for 
all $t$).  This is a Metropolis update within Gibbs; for 
each year $t$, 10 positions in the current permutation 
vector $\perm_t^c$ were randomly selected.  Then the values
of those 10 positions were randomly shuffled, with the 41 
other positions remaining fixed.  This modified permutation
vector is the proposal $\perm_t^*$.  Since the proposal 
distribution is symmetric, and the likelihood does not 
depend on the permutation, this proposal is accepted with 
probability 
\[ \min \left(1, \frac{p_\text{sp}(\partition_t \mid 
\baselinepart, \shrink, \perm_t^*, \psi, \baselinedist) }
{p_\text{sp}(\partition_t \mid \baselinepart, \shrink, 
\perm_t^c, \psi, \baselinedist)} \right). \]
This Metropolis update procedure is repeated 9 more times
to improve mixing.  The specific choice of shuffling 10 
positions and 10 Metropolis updates are tuning
details, which can be increased or decreased as appropriate.
$\perm_t$ is initialized at the natural permutation, 
$(1,2,\ldots,n)$.

\item $\bm{\beta}^*_{ct}$: The prior for each state and 
year is $\bm{\beta}_{it} \sim 
\text{N}\left(\bm{\mu}_{\beta}, \tau_{\beta} \bm{I} 
\right)$, \textit{iid} for all $i$ and $t$.  In this model,
we set $\bm{\mu}_{\beta}=(1.46,0.15,0.24,0.41)^{'}$ and 
$\tau_{\beta}=100$.  This step is a Gibbs update.  For each
combination of $c$ and $t$, draw an update as follows:
\[ \bm{\beta}^*_{ct} \mid \cdot \ \sim 
\text{N}\left(\left[\tau_t \bm{X}^{'}_{ct} \bm{X}_{ct} 
+\tau_{\beta} \bm{I}\right]^{-1}\left[ \tau_t 
\bm{X}^{'}_{ct} \left(\bm{y}_{ct} - 
\bm{Z}_{ct}\bm{\gamma}_t\right) + \tau_{\beta} \left( 
\bm{\mu}_{\beta} - \bm{Z}_{ct}\bm{\gamma}_t \right)\right],
\tau_t \bm{X}^{'}_{ct} \bm{X}_{ct} +\tau_{\beta} \bm{I} 
\right). \]
We note here that the second argument in the normal 
distribution is a precision (not a variance or standard 
deviation).  Additionally, the ``$\cdot$" in the 
conditioning symbol is used to denote all other parameters, 
without enumerating them.
Since $\partition_t$ is initialized at the census regions
(i.e., it has four clusters), the four corresponding 
$\bm{\beta}^*_{ct}$'s for each year are initialized by 
drawing from their independent 
$\text{N}(\bm{\mu}_{\beta}, \tau_{\beta} \bm{I} 
)$ priors.

\item $\bm{\gamma}_t$: The prior for each year is 
$\bm{\gamma}_{t} \sim \text{N}\left(\bm{0}, \tau_{\gamma}
\bm{I} \right)$, \textit{iid} for all $t$. In this model 
we set $\tau_{\gamma}=1$.  This MCMC step is a Gibbs 
update. For each year $t$, draw an update as follows:
\[ \bm{\gamma}_{t} \mid \cdot \ \sim  
\text{N}\left(\left[\tau_t \bm{Z}^{'}_t \bm{Z}_t +
\tau_{\gamma} \bm{I}\right]^{-1}\left[ \tau_t \bm{Z}^{'}_t 
\left(\bm{y}_t - 
\begin{bmatrix}
  \bm{X}_{1t}\bm{\beta}_{1t} \\
  \vdots \\
  \bm{X}_{nt}\bm{\beta}_{nt}
\end{bmatrix} \right) \right], \tau_t \bm{Z}^{'}_t 
\bm{Z}_t +\tau_{\gamma} \bm{I} \right).
\]
The value of $\bm{\gamma}_t$ is initialized by a random
draw from its prior $\text{N}\left(\bm{0}, \tau_{\gamma} 
\bm{I} \right)$, independently for each year.

\item $\tau_t$: The prior for each year is $\tau_t \sim 
\text{Gamma}(\alpha_{\tau},\beta_{\tau})$, \textit{iid} 
for all $t$. We set $\alpha_{\tau} = 
1/0.361^2$, and $\beta_{\tau}=1$.  This step is a Gibbs 
update. For each year $t$, draw an update as follows: 
\[ \tau_t \mid \cdot \ \sim \text{Gamma}\left(\alpha_{\tau}
+ \frac{1}{2} \sum_{i=1}^{n} n_{it}, \ \beta_{\tau} + 
\frac{1}{2} \sum_{i=1}^{n}  \left(\bm{y}_{it} - 
\bm{X}_{it}\bm{\beta}_{it} - \bm{Z}_{it}\bm{\gamma}_{t} 
\right)^{'} \left( \bm{y}_{it} - \bm{X}_{it}\bm{\beta}_{it}
- \bm{Z}_{it}\bm{\gamma}_{t} \right) \right). \]
We use the rate parameterization of the 
gamma distribution.  The value of $\tau_t$ is initialized 
by a random draw from its prior, independently for each 
year.

\item $\shrink$: The difference in the two independent 
models is solely the choice of $\shrink$.  In the 
independent model with common shrinkage 
$\shrink = (5,\ldots,5)$, i.e., all states have a shrinkage
value of 5.  For the independent model with idiosyncratic 
shrinkage, each state had a shrinkage value set to 5 except 
for Maryland, Delaware, District of Columbia, Montana, 
North Dakota, and South Dakota, whose shrinkage values were 
set to 1.  Since these vectors are fixed, no MCMC update is 
needed.

\item $\psi$: In the independent model, the grit parameter
is fixed at 0.02.  Therefore, no updates are needed. 
\end{enumerate}

This model allows for each year to be run in parallel.

\subsection{Hierarchical Model} \label{sec:hierMCMC}

This section provides the MCMC details for the hierarchical model in Section
\ref{sec:hierarchical}.  This model is similar to the independent
model described in the previous section. The primary difference is that each
state now has a common anchor partition, which itself has a prior distribution.
The following list describes how the parameters are updated and initialized.

\begin{enumerate}

\item $\partition_t$: The prior for each $\partition_t \mid
\baselinepart,\shrink,\perm_t,\psi,\baselinedist$ is an 
independent $\text{SP}
(\baselinepart,\shrink,\perm_t,\psi,\baselinedist)$.  For 
each year, update the partition according to the steps in 
Section \ref{sec:modAlgo8}.  The value of $\partition_t$ is 
initialized by randomly (and uniformly) drawing a value 
from $\{1,2,3,4,5\}$ for each state.  The $\partition_t$ for
each year $t$, is given the same initial value. 
$\baselinedist$ is the CRP(1), and the other parameters are 
described below.

\item $\baselinepart$: In this model, the anchor is common 
across all years, which permits the borrowing of strength 
between all $\partition_t$.  The prior is $\baselinepart 
\sim \text{SP}(\bm{\rho}_{\text{r}},\shrink_{0},
\perm_{0},
\psi_{0}, \baselinedist)$, where $\baselinedist$ is the 
CRP(1).  In the following sub-steps we show how 
$\baselinepart$ and its associated hyperparameters are 
updated.
\begin{enumerate}[a)]
\item To update $\baselinepart$, we use a one-at-a-time 
Gibbs update based on Algorithm 8 in \cite{neal2000markov}.
For all $i = 1,\ldots,n$ repeat the following steps:
\begin{enumerate}[i)]
    \item Let $q_{\baselinepart}$ be equal to the number of
    clusters in the current value of $\baselinepart$, and
    let $\baselinepart^*$ be identical to $\baselinepart$,
    except that $\baselineitem_{\permitem_{0 i}}^* = c$. 
    Then, for $c=1,\ldots,q_{\baselinepart}$, calculate the 
    following quantity: 
    \begin{align} \label{eq:algo8probs1mu}
    \begin{split}
	\left( 
    \prod_{t=1}^{T} p_\text{sp}(\partition_t \mid 
    \baselinepart^*, \shrink, \perm_t, \psi, \baselinedist)
    \right) \times
    &p_\text{sp}(\baselinepart^* \mid
    \bm{\rho}_{\text{r}},\shrink_{0},\perm_{0},
    \psi_{0},\baselinedist).
    \end{split}
    \end{align} 
    \item If the ${\permitem_{0i}}^{\text{th}}$ item of 
    $\baselinepart$ is not currently in a singleton 
    cluster, set $\baselinepart^*$ to be equal to 
    $\baselinepart$ except that 
    $\baselineitem_{\permitem_{0 i}}^* = q_{\baselinepart}
    +1$. Then calculate (\ref{eq:algo8probs1mu}) again.  
    If the ${\permitem_{0i}}^{\text{th}}$ item of 
    $\baselinepart$ is in a singleton cluster, omit this 
    step.
    \item With probabilities proportional to the quantities
    calculated in the previous two steps, randomly allocate
	the ${\permitem_{0i}}^{\text{th}}$ item to a cluster.
\end{enumerate}

    \item $\perm_{0}$: This is the same as described in 
    Section \ref{sec:indMCMC}, except that only 5 positions
    in the proposed permutation vector are shuffled. 
    \item $\shrink_{0}$: $\shrink_{0}$ is defined as 
    $\shrinkitem_{0} \, \times \, (a_{\permitem_{01}}, 
    a_{\permitem_{02}}, \ldots, a_{\permitem_{0n}})$, where
    $a_{\permitem_{0i}}$ is equal to 1, except for 
    Maryland, Delaware, District of Columbia, Montana, 
    North Dakota, and South Dakota whose $a_{\permitem_{0i}}$ values are set to $0.2$. 
    The prior for $\shrinkitem_{0}$ is a $\text{Gamma}
    (5,1)$ and $\shrinkitem_{0}$ is updated via slice 
    sampling with step size of $\sqrt{5}$.  It is 
    initialized with a random draw from its prior.
    \item $\psi_{0}$: The prior for $\psi_{0}$ is a 
    uniform on the unit interval.  This parameter is updated using the 
    slice sampler with a step size of $1/\sqrt{12}$. The 
    initial value of $\psi_{0}$ is randomly drawn from its 
    prior.
\end{enumerate}

\item $\perm_t$: This is the same as described in Section 
\ref{sec:indMCMC}, except that only 5 positions in the 
proposed permutation vector are shuffled, to improve  
acceptance rates.
\item $\bm{\beta}^*_{ct}$: This is the same as described in 
Section \ref{sec:indMCMC}.
\item $\bm{\gamma}_t$: This is the same as described in 
Section \ref{sec:indMCMC}.
\item $\tau_t$: This is the same as described in Section 
\ref{sec:indMCMC}.
\item $\shrink$: Let $\shrink = \shrinkitem \times (1,\ldots,1)$
and assume $\shrinkitem \sim \text{Gamma}(5,1)$.
Update using a slice sampler with a step size of $\sqrt{5}$.
The value of $\shrinkitem$ is initialized by taking a draw
from its prior.
\item $\psi$: The prior for $\psi$ is a uniform on the unit interval. 
This parameter is also updated using the slice sampler, 
with a step size of $1/\sqrt{12}$. The initial value of 
$\psi$ is randomly drawn from its prior.
\end{enumerate}

\subsection{Temporally-Dependent Model} 
\label{sec:temporalMCMC}
The temporally-dependent model, as implemented in Section 
\ref{sec:temporal}, has many common aspects of the previous
models.  However, in this model, we do not define a 
$\baselinepart$ (except at year $t=1$) since we use the
previous year's partition $\partition_{t-1}$ as the anchor 
for the current year.  In this model all baseline 
distributions $\baselinedist$ are the CRP(1).

\begin{enumerate}

\item $\partition_1$:  The prior for $\partition_1 \mid 
\bm{\rho}_{\text{r}},\shrink_1,\perm_1,\psi_1,\baselinedist$
is an independent $\text{SP}
(\bm{\rho}_{\text{r}},\shrink_1,\perm_1,\psi_1,\baselinedist
)$.

\begin{enumerate}[a)]
\item To update $\partition_1$, we again use a one-at-a-time
Gibbs update based on Neal's Algorithm 8.
For all $i = 1,\ldots,n$, repeat the following steps:
\begin{enumerate}[i)]
    \item Let $q_{\partition_1}$ be the number of clusters 
    in current version of $\partition_1$, and define  
    $\partition_1^*$ to be identical to $\partition_1$, 
    except that $\partitionitem_{1\permitem_{1 i}}^* = c$. 
    Then, for $c=1,\ldots,q_{\partition_1}$, compute the 
    following quantity:  
    \begin{equation} \label{eq:algo8probs1pi1}
    p(\bm{y}_{i1}\mid \bm{\beta}^*_{c1},\bm{\gamma}_1,\tau_1) \times 
    p_\text{sp}(\partition_1^* \mid  \bm{\rho}_{\text{r}},\shrink_{1},\perm_{1},\psi_{1}, \baselinedist), \times
    p_\text{sp}(\partition_2 \mid \partition_1^*, \shrink,\perm_2, \psi, \baselinedist).
    \end{equation} 
    \item If the ${\permitem_{1i}}^{\text{th}}$ item of 
    $\partition_1$ is not currently in a singleton cluster,
    set $\partition_1^*$ to be equal to $\partition_1$ 
    except that $\partitionitem_{1\permitem_{1 i}}^* = 
    q_{\partition_1} +1$. Then calculate 
    (\ref{eq:algo8probs1pi1}) again.  If the 
    ${\permitem_{1i}}^{\text{th}}$ item of $\partition_1$
    is in a singleton cluster, omit this step.
    \item With probabilities proportional to the quantities
    calculated in the previous two steps, randomly select 
    the cluster to which the 
    ${\permitem_{1i}}^{\text{th}}$ item of $\partition_1$ 
    is reassigned.
\end{enumerate}
    \item $\perm_{1}$: This is the same as described in 
    Section \ref{sec:indMCMC}, except that only 5 positions 
    in the proposed permutation vector are shuffled. 
    \item $\shrink_{1}$: $\shrink_{1}$ is defined as 
    $\shrinkitem_{1} \, \times \, (a_{\permitem_{11}}, 
    a_{\permitem_{12}}, \ldots, a_{\permitem_{1n}})$, where 
    $a_{\permitem_{1i}}$ is equal to 1, except for 
    Maryland, Delaware, District of Columbia, Montana, 
    North Dakota, and South Dakota whose $a_{\permitem_{1i}}$ values are $0.2$. 
    The prior for $\shrinkitem_{1}$ is a $\text{Gamma}
    (5,1)$.  $\shrinkitem_{1}$ is updated via slice 
    sampling, with step size of $\sqrt{5}$, and it is 
    initialized with a random draw from its prior.
    \item $\psi_{1}$: The prior for $\psi_{1}$ is a 
    uniform on the unit interval.  This parameter is updated using the 
    slice sampler with a step size of $1/\sqrt{12}$. The 
    initial value of $\psi_{1}$ is randomly drawn from its
    prior.
\end{enumerate}

\item $\partition_t$ for $t=2,\ldots,T$: The prior for 
$\partition_t \mid \partition_{t-
1},\shrink,\perm_t,\psi,\baselinedist$ is an independent 
$\text{SP}(\partition_{t-
1},\shrink,\perm_t,\psi,\baselinedist)$. 
This is a one-at-a-time Gibbs update based on Neal's Algorithm 8.
The value of each $\partition_t$
(including $\partition_1$), is initialized the same as in 
Section \ref{sec:hierMCMC}.  For each year $t=2,\ldots,T$ 
and for all $i = 1,\ldots,n$, repeat the following steps:  

 \begin{enumerate}[a)]
    \item Let $q_{\partition_t}$ to equal to the number of 
    clusters in the current value of $\partition_t$.  
    Additionally, set $\partition_t^*$ to be identical to 
    $\partition_t$, with the exception that 
    $\partitionitem_{t\permitem_{t i}}^* = c$.  Then for 
    $c=1,\ldots,q_{\partition_t}$, compute the following 
    quantity:
    \begin{equation} \label{eq:algo8probs1pit}
    p(\bm{y}_{it}\mid  
    \bm{\beta}^*_{ct},\bm{\gamma}_t,\tau_t)  \times
    p_\text{sp}(\partition_t^* \mid \partition_{t-
    1},\shrink,\perm_{t},\psi,\baselinedist)
    \times p_\text{sp}(\partition_{t+1} \mid 
    \partition_t^*, \shrink, \perm_{t+1}, \psi, 
    \baselinedist).
    \end{equation}
    Note: When $t=T$ the last term should be omitted from
    (\ref{eq:algo8probs1pit}).  
    \item If the ${\permitem_{ti}}^{\text{th}}$ item of 
    $\partition_t$ is not currently in a singleton cluster,
    set $\partition_t^*$ to be equal to $\partition_t$, 
    with the exception that 
    $\partitionitem_{t\permitem_{t i}}^* = 
    q_{\partition_t} +1$. Then calculate 
    (\ref{eq:algo8probs1pit}) again.  If the 
    ${\permitem_{ti}}^{\text{th}}$ item of $\partition_t$ 
    is in a singleton cluster, omit this step.
    \item With probabilities proportional to the quantities
    calculated in the previous two steps, randomly select
    the cluster to which the 
    ${\permitem_{ti}}^{\text{th}}$ item of $\partition_t$ 
    is reassigned.
\end{enumerate}

\item $\perm_t$: This is the same as described in Section
\ref{sec:hierMCMC}, except the acceptance probability is 
now: 
\[ \min \left(1, \frac{p_\text{sp}(\partition_t \mid 
\partition_{t-1}, \shrink, \perm_t^*, \psi, \baselinedist) }
{p_\text{sp}(\partition_t \mid \partition_{t-1}, \shrink, 
\perm_t^c, \psi, \baselinedist)} \right). \]
\item $\bm{\beta}^*_{ct}$: This is the same as described in
Section \ref{sec:indMCMC}.
\item $\bm{\gamma}_t$: This is the same as described in 
Section \ref{sec:indMCMC}.
\item $\tau_t$: This is the same as described in Section
\ref{sec:indMCMC}.
\item $\shrink$: This is the same as described in Section 
\ref{sec:hierMCMC}.
\item $\psi$: This is the same as described in Section
\ref{sec:hierMCMC}.
\end{enumerate}
\newpage

\section{US Census Regions}
\label{sec:censusregions}

\begin{figure}[h]
	\includegraphics[width=0.70\textwidth]{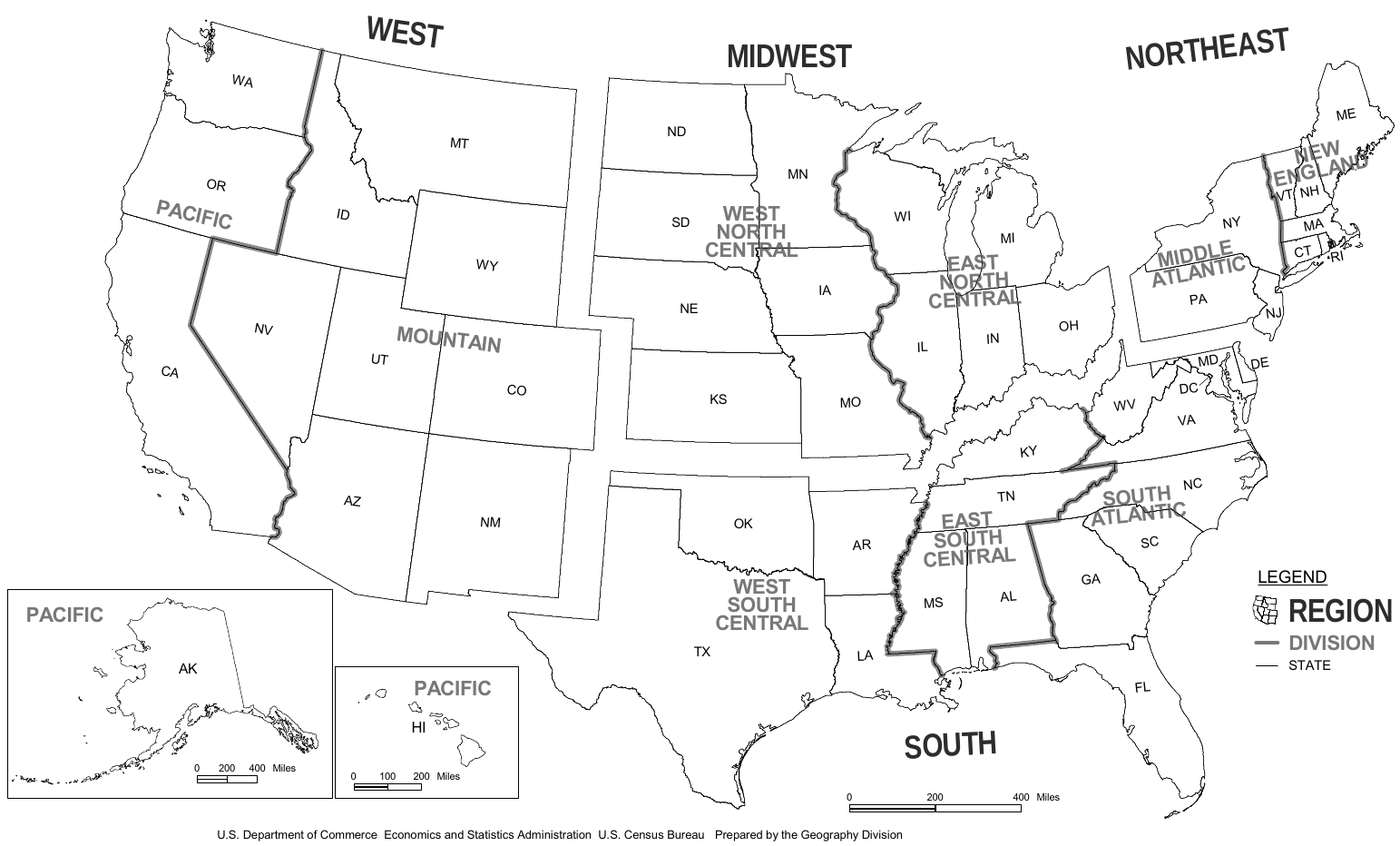}
	\caption{The four regions defined by U.S. Census Bureau are used as a prior anchor partition 
            in the education regression example in Subsection \ref{sec:single}
		in which the regression coefficients for the 50 states and the District of Columbia 
            are clustered.}
	\label{fig:regions}
\end{figure}

\newpage{}
\section{Cluster Estimation}
\label{sec:clusterestimation}

To study cluster estimation, we use the same
model as Section \ref{sec:single} and only consider one year (so we drop $t$
in the notation). The values of the model parameters in the simulation were based
on OLS estimates from the IPUMS data in Section \ref{sec:AppRTE}, using a model without state-specific
intercepts and slopes.  For 50 of the simulated datasets, each of the $n = 51$
simulated states had exactly $m=100$ observations.  Another 50 datasets were
simulated with $m=400$ observations. Responses for each dataset were generated
from the sampling model as follows.  The regression coefficient vector $
\bm{\gamma}$ for the non-education covariates was set at the OLS estimates
and the covariate values were sampled from the actual data for each state.
The regions partition $\bm{\rho}_\text{r}$ was used to generate data from
four clusters and the education coefficients $\bm{\beta}_c^*$ for each cluster
were deviations from the OLS estimates.  The education covariate values
were sampled from the actual data for each state.  The error precision for the
simulated data was obtained from the OLS estimate.

The posterior expected Rand index $\text{E}(\,\text{RI}(\partition,\bm{\rho}_\text{r})
\mid \bm{y})$ is displayed on the left hand side of
Figure \ref{fig:simulation}. Specifically, we assume $\partition \sim \text{SP}
(\bm{\mu}, \shrink, \perm, \psi, \text{JLP(1)})$ with $\shrink = \shrinkitem
\times (1, \ldots, 1)$ for $\shrinkitem = 0, 1, \ldots, 6$, $\perm$ having a
discrete uniform distribution, and $\psi$ having a $\text{Beta}(2,2)$ distribution (to match the prior elicitation in the next paragraph).
We let $\bm{\mu}$ equal either the regions partition $\bm{\rho}_\text{r}$
or the partition $\bm{\rho}_\text{s}$ obtained by randomly shuffling the region labels. When $\bm{\mu}
= \bm{\rho}_\text{r}$, i.e., the partition from which the data was generated, the
posterior expected Rand index improves with increasing shrinkage $\shrinkitem$.
Conversely, when assuming $\bm{\mu} = \bm{\rho}_\text{s}$, this mistake leads to
worsening posterior expected Rand index as the shrinkage $\shrinkitem$
increases. Also, note that the posterior expected Rand index improves with
the sample size $m$ increasing from 100 to 400.  The upshot is \estimationupshot{}

\begin{figure}[tb]
     \centering
     \begin{minipage}{0.49\textwidth}
         \centering
         \resizebox{0.99\linewidth}{!}{\input{img/simulation_1}}
     \end{minipage}
	 \hfill
     \begin{minipage}{0.49\textwidth}
         \centering
         \renewcommand{\arraystretch}{0.6}
         \vspace{-2ex}
         \resizebox{0.99\linewidth}{!}{\begin{tabular}{clcc}
             \toprule
              &  & \multicolumn{2}{c}{$\text{E}(\text{Rand Index} \mid \text{Data})$} \\
              &  & \multicolumn{2}{c}{w/ Anchor $\bm{\mu}$ Being...} \\
             \cmidrule(lr){3-4}
             $m$ & Distribution & Regions & Shuffled \\
             \midrule
             100 & SP w/ random $\shrinkitem$ and $\psi$   & (0.91, 0.94) & (0.68, 0.70) \\
             100 & LSP w/ random $\shrinkitem$             & (0.86, 0.88) & (0.68, 0.69) \\
             100 & CPP-VI w/ fixed $\shrinkitem$           & (0.85, 0.87) & (0.62, 0.64) \\
             100 & CPP-B w/ fixed $\shrinkitem$            & (0.93, 0.94) & (0.72, 0.73) \\
             \midrule
             400 & SP w/ random $\shrinkitem$ and $\psi$   & (0.96, 0.98) & (0.75, 0.76) \\
             400 & LSP w/ random $\shrinkitem$             & (0.93, 0.95) & (0.74, 0.75) \\
             400 & CPP-VI w/ fixed $\shrinkitem$           & (0.93, 0.94) & (0.74, 0.75) \\
             400 & CPP-B w/ fixed $\shrinkitem$            & (0.97, 0.98) & (0.77, 0.78) \\
             \bottomrule
             \end{tabular}}
     \end{minipage}
     \caption{Left: The posterior expected Rand index for the true partition $\bm{\rho}_\text{r}$ used to generate the data and models with the SP distribution, as a function of the shrinkage $\shrinkitem$, the assumed anchor $\bm{\mu}$, and the sample size $m$.
     Right: The posterior expected Rand index for the true partition $\bm{\rho}_\text{r}$ for various prior distributions, all of which have been calibrated to have the same prior expected Rand index as the SP distribution. 95\% confidence intervals are shown.}
  \label{fig:simulation}
\end{figure}

\begin{figure}[tb]
     \centering
     \begin{subfigure}[b]{0.32\textwidth}
         \centering
         \resizebox{0.98\linewidth}{!}{\includegraphics{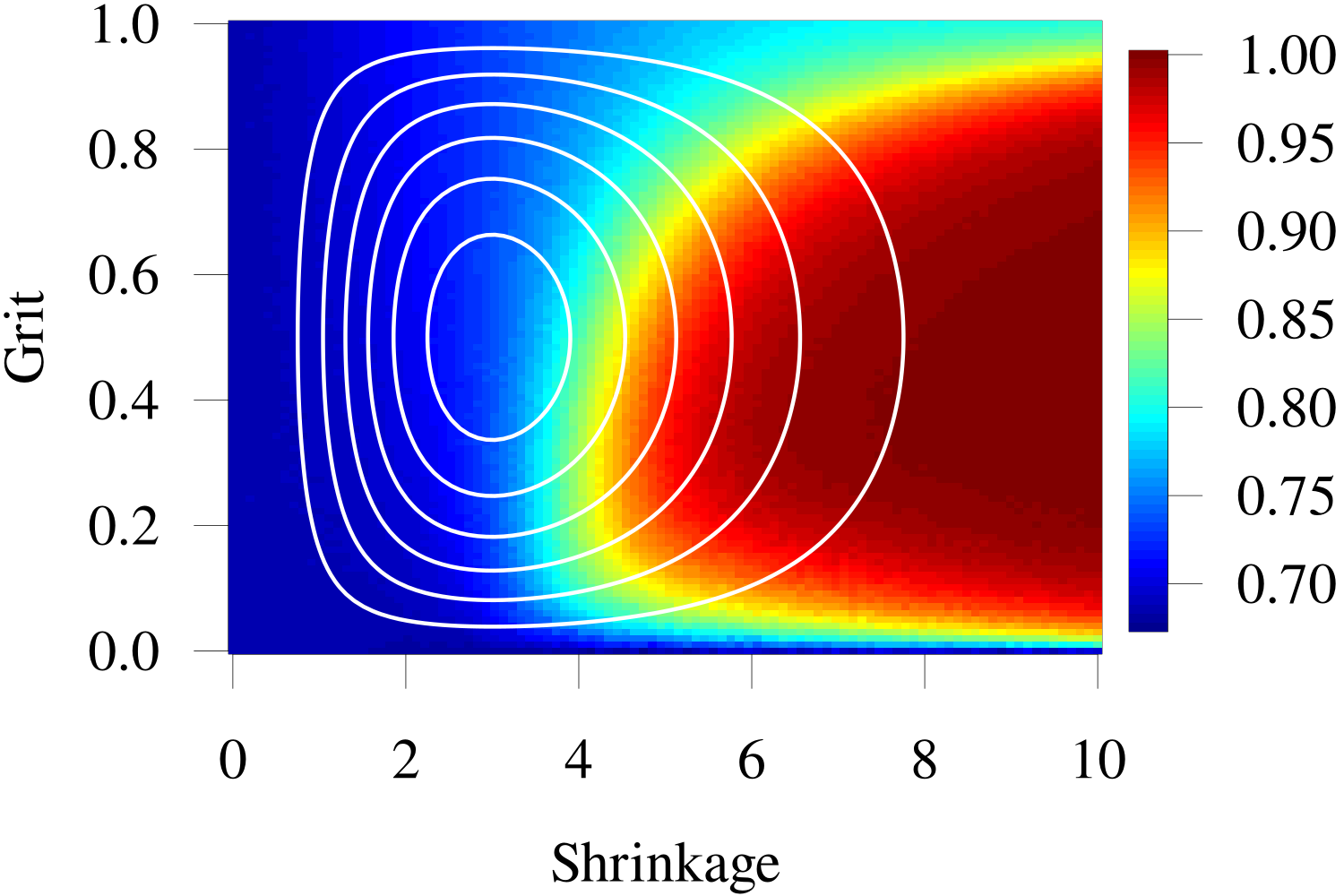}}
     \end{subfigure}
     \hfill
     \begin{subfigure}[b]{0.32\textwidth}
         \centering
         \resizebox{0.98\linewidth}{!}{\includegraphics{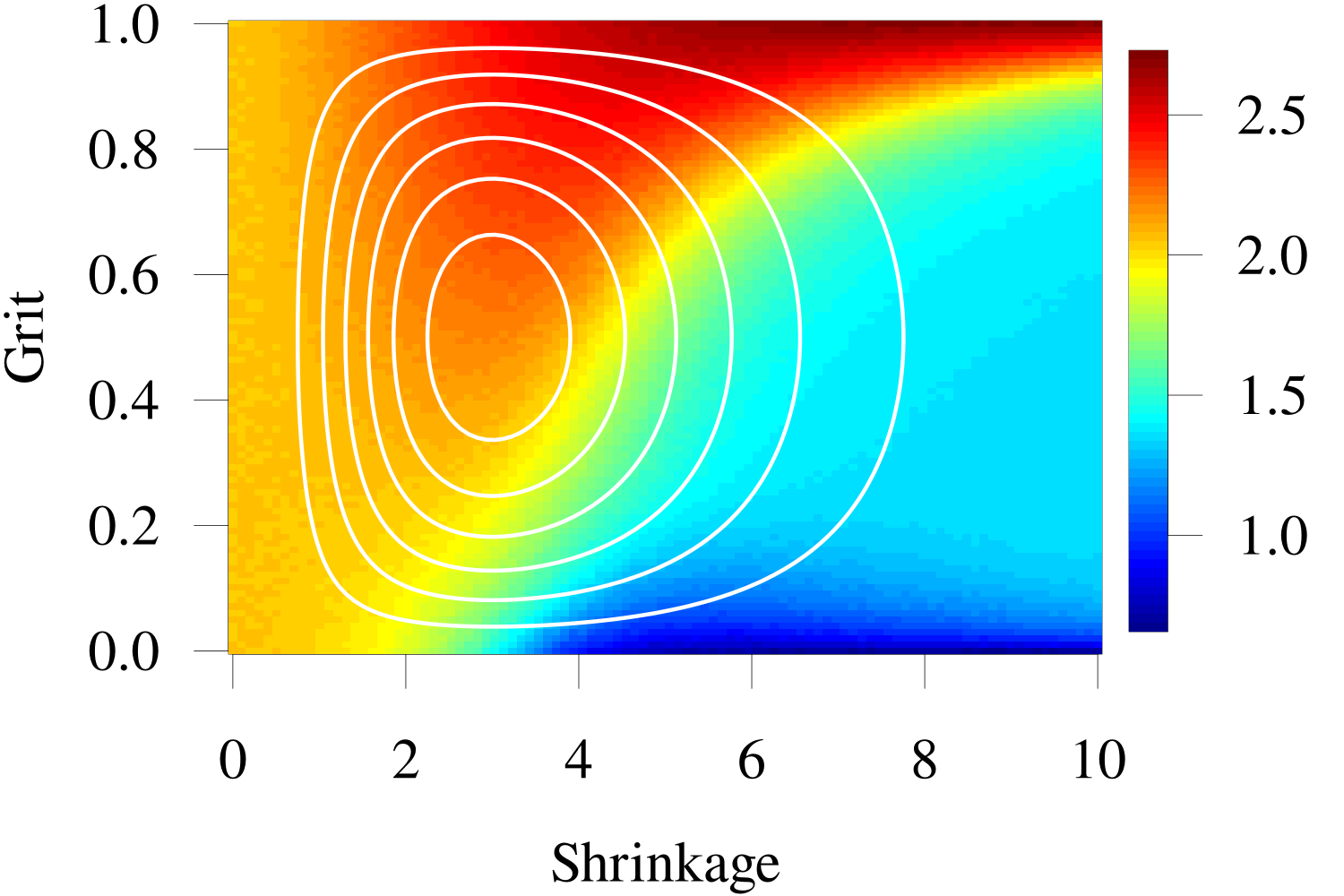}}
      \end{subfigure}
     \hfill
     \begin{subfigure}[b]{0.32\textwidth}
         \centering
         \resizebox{0.98\linewidth}{!}{\includegraphics{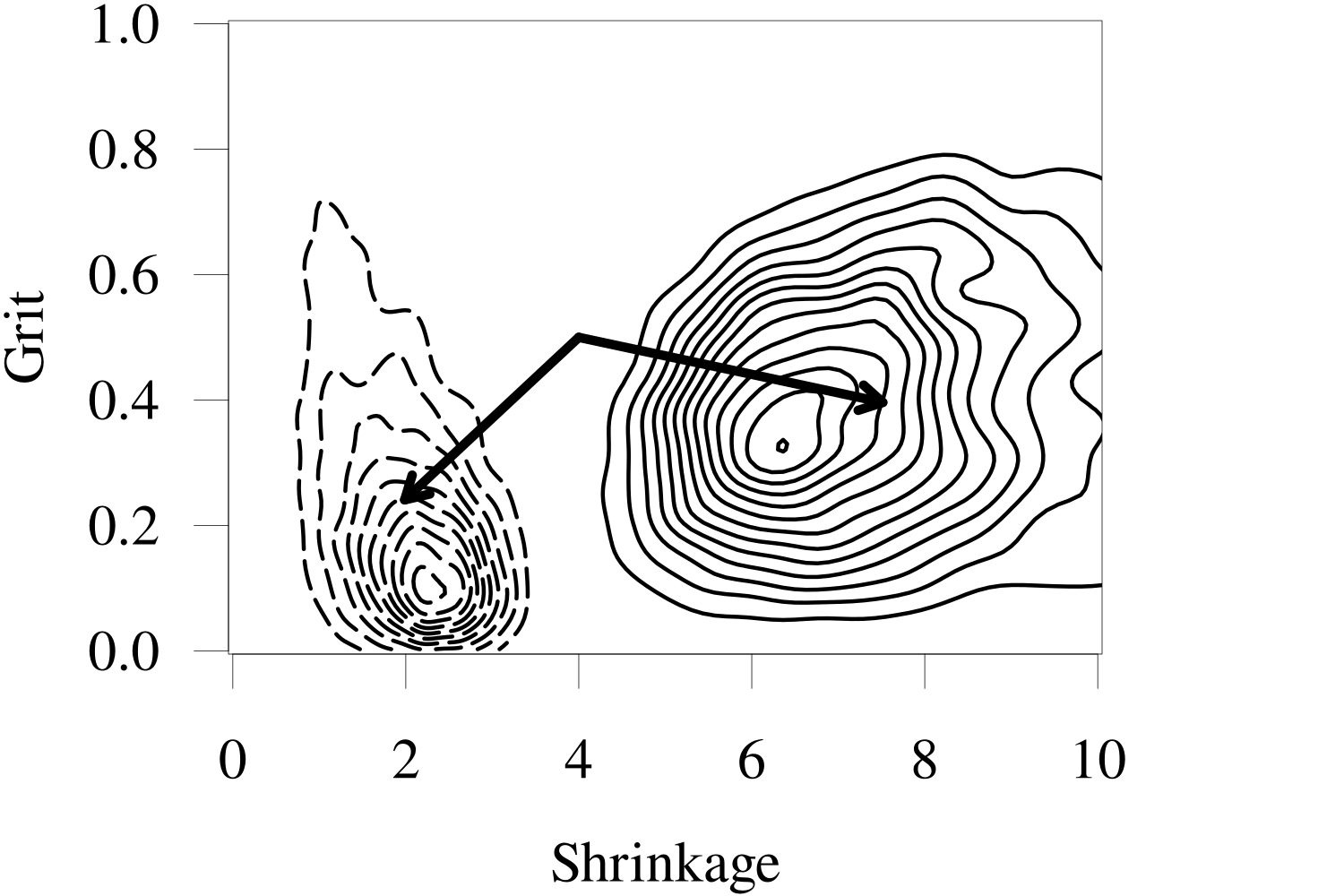}}
      \end{subfigure}
     \caption{Left: Prior density of the joint prior on the shrinkage $\shrinkitem$ and the grit $\psi$ in white contours, with the prior expected Rand index in color.  Middle: Same as left plot, except the prior expected entropy is in color.  Right: Posterior densities of the shrinkage $\shrinkitem$ and the grit $\psi$ for misspecified anchor $\bm{\rho}_s$ (dashed) and the correctly specified anchor $\bm{\rho}_r$ (solid).  The arrows indicate the movement of the prior mean to the posterior means.}
     \label{fig:simulation_shrinkage_grit}
\end{figure}

Before comparing the SP, LSP, and CPP distributions for partition estimation,
we first need to consider the prior on the shrinkage parameter $\shrinkitem$
and grit parameter $\psi$ and then how to make the LSP and CPP distributions
put the same amount of prior information on the anchor partition. We use the
prior elicitation method outlined in Section \ref{sec:simulateprior}. Consider
shrinkage values $\shrink = \shrinkitem \times (1, \dots, 1)$ from $\shrinkitem$ on a grid
ranging from 0 to 10 and grid of grit values $\psi$ from 0 to 1. For each
combination of $\shrinkitem$ and $\psi$ values, sample many partitions from
the SP distribution using uniformly-sampled permutations, a JLP(1) baseline
distribution, and the anchor set to $\bm{\rho}_\text{r}$ and then compute Monte
Carlo estimates of
$\text{E}(\text{RI}(\partition,\bm{\rho}_\text{r}))$ and $\text{E}(\text{Entropy}(\partition))$.
The left and middle
panels of Figure \ref{fig:simulation_shrinkage_grit} show heatmap plots of
the expected Rand index (left) and the expected entropy (middle).
Notice that the shrinkage $\shrinkitem$ has a strong influence on the expected
Rand index whereas the grit $\psi$ substantially changes the expected cluster
entropy. The plots
also show joint density contours for $\shrinkitem \sim \text{Gamma}(4,1)$ and $\psi \sim \text{Beta}(2,2)$.
We suggest this is a justifiable joint prior since the distribution has substantial density for
a wide range of values for the Rand index and entropy, but one could modify
the prior specification for $\shrinkitem$ and $\psi$ if the contours in the
heatmap plots do not align with prior expectations.
The right plot of Figure \ref{fig:simulation_shrinkage_grit}
\posteriorlearning{}

Now we compare the SP, LSP, and CPP distributions for partition estimation.
Recall that LSP reduces to $\text{JLP}(1)$ when shrinkage $\shrinkitem = 0$, 
so we use $\text{JLP}(1)$ as the baseline distribution
for SP and CPP for the sake of comparison.
We want to compare the models for a finite shrinkage using both a correctly
specified anchor $\bm{\mu} = \bm{\rho}_\text{r}$ and a misspecified anchor $\bm{\mu} =
\bm{\rho}_\text{s}$. Since the SP, LSP, and CPP have different formulations, each
distribution must be calibrated such that it imbues the same degree of belief in the anchor $\bm{\mu}$.  To this end,
we computed the prior expected Rand index $\text{E}(\text{RI}(\partition,\bm{\mu}))$
for the SP distribution with 
$\shrinkitem \sim \text{Gamma}(4,1)$ and $\psi \sim \text{Beta}(2,2)$
under both anchors
$\bm{\mu} = \bm{\rho}_\text{r}$ and
$\bm{\mu} = \bm{\rho}_\text{s}$.
We then found, for each anchor, a $u$ such that the LSP with a $\text{Gamma}(u,1)$ prior on its shrinkage
parameter led to the same prior expected Rand index as the SP distribution for that anchor.  We did
likewise for the CPP with Binder loss and VI loss, except we found fixed shrinkage values
since the CPP cannot accommodate priors on hyperparameters.
With these calibrated priors, we fit the model for our 100 simulated datasets and
the table on the right hand side of
Figure \ref{fig:simulation} shows 95\% confidence intervals on the Monte Carlo estimates of
$\text{E}(\,\text{RI}(\partition,\bm{\rho}_\text{r}) \mid \bm{y})$.
Note that, for both $m=100$ and $m=400$, the SP
distribution performs about as well as the CPP with Binder loss when the anchor is the
truth (i.e., $\bm{\mu} = \bm{\rho}_\text{r}$) and better than LSP and CPP with VI loss.
This is interesting
because the CPP with Binder loss performed poorly in Section \ref{sec:single}, yet 
performs well here.
When the anchor is mistaken (i.e., $\bm{\mu} = \bm{\rho}_s$),
most of the 95\% confidence intervals overlap.
We also found (but do not show) that the SP distribution with $\bm{\mu} = \bm{\rho}_\text{r}$
is the best in terms of 
$\text{E}(\,\text{VI}(\partition,\bm{\rho}_\text{r}) \mid \bm{y})$ and has non-overlapping confidence intervals with
the LSP and CPP with both Binder and VI losses.
We \estimationcomparison{}

\end{appendices}

\end{document}